\documentclass[times]{qjrms4}
\usepackage{url}
\usepackage[dvipsnames]{xcolor}
 
\newcommand{\beqn}{\begin{equation}}
\newcommand{\eeqn}{\end{equation}}
\newcommand{\beqa}{\begin{eqnarray}}
\newcommand{\eeqa}{\end{eqnarray}}
\newcommand{\beqanonum}{\begin{eqnarray*}}
\newcommand{\eeqanonum}{\end{eqnarray*}}
\newcommand{\beqnonum}{\begin{equation*}}
\newcommand{\eeqnonum}{\end{equation*}}
\newcommand{\n}{\nonumber}

\newcommand{\bbf}{\begin{bf}}
\newcommand{\ebf}{\end{bf}}
\newcommand{\eqnref}[1]{(\ref{#1})}

\newcommand{\inverse}{^{-1}}

\newcommand{\meter}{\ensuremath{\mathrm{m}}}

\newcommand{\km}{\ensuremath{\mathrm{km}}}

\newcommand{\zct}{\ensuremath{z_{\rm ct}}}
\newcommand{\enet}{\ensuremath{\epsilon_{\rm net}}}
\newcommand{\Md}{\ensuremath{M_{\rm d}}}
\newcommand{\rth}{\ensuremath{r_{\rm th}}}
\newcommand{\rhoth}{\ensuremath{\rho'_{\rm th}}}
\renewcommand{\Re}{\ensuremath{{{\rm Re}}}}

\newcommand\Beq{\begin{eqnarray}}
\newcommand\Eeq{\end{eqnarray}}

\renewcommand{\vec}[1]{\boldsymbol{#1}}

\newcommand{\note}[1]{#1}

\newcommand{\newnote}[1]{#1}

\begin{document}

\runningheads{D.~Lecoanet \& N.~Jeevanjee}{Entrainment in Resolved Thermals}

\title{\newnote{Entrainment in Resolved, Dry Thermals}}

\author{D.~Lecoanet\affil{a,b} and N.~Jeevanjee\affil{c} }

\address{\affilnum{a}Princeton Center for Theoretical Science, Princeton, NJ 08544, USA \\
\affilnum{b}Department of Astrophysical Sciences, Princeton, NJ 08544, USA \\
\affilnum{c}Department of Geosciences, Princeton, NJ 08544, USA}

\begin{abstract}
Entrainment in cumulus convection remains ill-understood and difficult to quantify. For instance, entrainment is widely believed to be a fundamentally turbulent process, even though Turner (1957) pointed out that dry thermals entrain primarily because of buoyancy  (via a dynamical constraint requiring an increase in radius $\vec{r}$), rather than turbulence. Furthermore, entrainment has been postulated to obey a $1/\vec{r}$ scaling, but this scaling has not been firmly established.\\ Here, we study the classic case of dry thermals in a neutrally stratified environment using fully resolved direct numerical simulation. We combine this with a thermal tracking algorithm which defines a control volume for the thermal at each time, allowing us to directly measure entrainment. We test Turner's argument by varying the Reynolds number $\vec{\Re}$ of our thermals between laminar ($\vec{\Re\approx 600}$) and turbulent ($\vec{\Re\approx 6\,000}$) regimes,  finding only a 20\% variation in entrainment rate $\vec{\epsilon}$,  supporting the claim that turbulence is not necessary for entrainment.  We also directly verify the postulated $\vec{\epsilon \sim 1/r}$ scaling law. 

\end{abstract}

%\runningheads{D.~Lecoanet \& N.~Jeevanjee}{Entrainment in Resolved Thermals}

%\keywords{Convection; Large eddy and turbulence modeling; Theory; Turbulence; Atmosphere; Boundary layer; Local or boundary layer scale}

\maketitle

\section{Introduction}\label{sec:intro}
The rate at which cumulus clouds mix with their environment, or entrain, has long been known to be central to their dynamics
\citep{simpson1983a,cotton1975,simpson1965,stommel1947}. This led to a large number of studies, particularly in the early days, focused on the entrainment and dynamics of discrete, transient, convecting `thermals', believed to be the fundamental unit of convection \citep[see the review by][and references therein and given below]{yano2014a}.  With the advent of large-scale numerical modeling and convective parameterization schemes, however, attention shifted to describing the average entrainment of an \emph{ensemble} of convecting clouds,  often conveniently modeled as one or more continuous, steady-state, entraining `plumes' \citep{yano2014a,derooy2013}. While such plume convection schemes are ubiquitous in global climate models, they lie in tension with the wealth of evidence that cumulus clouds are actually composed of discrete thermals \citep[][]{romps2015, sherwood2013, heus2009, damiani2006,blyth2005,zhao2005a,carpenter1998,miller1983,saunders1961,malkus1955a, scorer1953}. Furthermore, single plume schemes suffer from an `entrainment paradox' in which no optimal  entrainment  rate exists \citep[e.g.][]{sherwood2013,mapes2011}. This leads to  uncertainties in the parameterization of entrainment, which turn out to be some of the largest contributors to uncertainties in climate sensitivity \citep{zhao2014,klocke2011, murphy2004}. 

Given the uncertainties in plume convection schemes and their tenuous connection to cumulus phenomenology, it seems worthwhile to turn back to thermals as a basis for understanding clouds and building parameterization schemes, as suggested by \cite{sherwood2013} \citep[see also][for a recent effort to reconcile the thermal and plume pictures]{morrison2017}. As with plume models, however, entrainment rates are key. The parameter of interest is typically the fractional gross entrainment rate $\epsilon$, defined to be the fraction of a parcel's volume (or mass, assuming small horizontal variations in density)  that it entrains per unit vertical distance traveled, in units of $\meter\inverse$.  A long-standing, widely used assumption regarding $\epsilon$, sometimes known as the \emph{entrainment assumption}, is that for a  thermal of radius $r$,
\beqn
	\epsilon  \ = \  e/r
	\label{ent1}
\eeqn
 \citep[][]{johari1992, turner1986, simpson1983b, simpson1969, turner1962, levine1959, morton1956}. The number $e$ is a constant which we refer to here as the \emph{entrainment efficiency}. The entrainment assumption is also widely applied in plume models, where it is sometimes formulated in terms of a `inflow velocity' \citep[e.g.][]{turner1986,morton1956}, but this is equivalent to \eqnref{ent1}. Importantly, in virtually all these cases,  entrainment is thought to arise from turbulent mixing at the interface between buoyant fluid and the environment. 

The $1/r$ form of the entrainment assumption \eqnref{ent1}  is a plausible inference from basic dimensional analysis, and has been indirectly verified in laboratory experiments.  In particular, `dry' thermals \note{propagating through a neutrally-stratified (`neutral') ambient} with no phase change or buoyancy source were studied by  \cite{richards1961}, \cite{scorer1957},  and \cite{morton1956} (hereafter R61, S57, and  MTT56 respectively);  these authors wrote down theories \newnote{based on} \eqnref{ent1}, and found their experiments in broad agreement with those theories. 

This work on idealized dry thermals \note{in a neutral ambient}\footnote{\newnote{These experimental works, and our current study, assume the fluid is Boussinesq, i.e., the background density is assumed to be nearly constant. Thus, a buoyant fluid parcel which does not entrain or detrain would maintain constant volume as it rises. We thus neglect effects from atmospheric density stratification which may be included in, e.g., the anelastic approximation.}} provides a foundation on which a thorough and fundamental understanding of cumulus thermals might be built, presumably in the hierarchical manner often adopted in climate modeling \citep{jeevanjee2017a, held2005}. 
%Though such dry laboratory thermals differ significantly from their moist atmospheric counterparts (which experience internal heating and cooling due to condensation and evaporation), it seems perhaps necessary to fully understand the dry case before attempting the moist case, since the most intractable phenomena of the latter (i.e. mixing) is present in the former. 
Despite the simplicity of this case, however,  questions remain.  Perhaps most surprisingly, the well-cited work of \cite{turner1957} suggests that entrainment in thermals is \emph{not} fundamentally turbulent, but rather arises from a \newnote{relationship} between the thermal's impulse $P$, circulation $K$, and radius $r$:
\beqn
	 P \ = \ \pi \rho K r^2 \ .
	 \label{turner1}
\eeqn
Here the circulation $K$ is computed along a roughly semi-circular circuit passing through the center of the thermal's vortex ring and then returning through the ambient fluid. If the density anomaly is negligible along this circuit, then the circulation should be constant \citep{fohl1967}. This conclusion is supported by experiments \citep{zhao2013}.

%\note{The momentum changes due to the volume integral of the buoyancy force, as well as the surface integrals of the pressure perturbation and viscous stress along the outer circuit. If the outer circuit is taken sufficiently outside the thermal, the pressure perturbation and viscous stresses can also be neglected, and thus $dP/dt > 0$ for a buoyant thermal.} 

As for the impulse $P$, it can be regarded as a generalized momentum of the fluid which can only be changed by external forces, such as gravity, and which is unaffected by internal forces such as pressure or viscosity \citep[e.g.][]{shivamoggi2010,batchelor2000}. In the presence of positive buoyancy and no other external forces, then, we must have $dP/dt>0$. Following \cite{turner1957}, we can then combine this with constant $K$  and  \eqnref{turner1} to deduce that
\beqn
	\frac{d (r^2)}{dt} =  \frac{1}{\pi\rho K} \frac{dP}{dt} \ > 0\ ,
	\label{turner2}
\eeqn
i.e. \emph{\note{buoyant} thermals must entrain}. \note{(Similarly, for a vortex ring of descending, negatively buoyant fluid, both $dP/dt<0$ and $K<0$, so the radius also increases.)} \note{Although this argument implicitly neglects the effects of viscosity, it does not rely on turbulent fluctuations. Thus, it predicts that laminar and turbulent thermals should entrain in the same way. Physically, this is because a vortex ring of a specified size and circulation can only translate at a given velocity; if the velocity changes, then the radius must change.} This implication is quite surprising in light of the vast literature attributing entrainment in thermals to turbulence, but so far has received little attention.

Another question is the veracity of the $1/r$ scaling in \eqnref{ent1}. The laboratory measurements of R61, S57, MTT56, and others  are consistent with such a scaling but have not verified it directly, and previous numerical investigations seeking a direct confirmation have yielded mixed results \citep[e.g.][]{hernandez2018,dawe2013, stirling2012}.  Finally,  S57 and others  studying the neutrally stratified case claim that detrainment is negligible, though this is never quantified and a detrained wake is clearly visible in photographs of the experiments. Thus,  there are still several open questions:
 
\begin{enumerate}
	\item Does entrainment depend strongly or weakly on \Re? \label{Q_Re}
	\item Does directly measured entrainment indeed obey the $1/r$ scaling \eqnref{ent1}? \label{Q_ent}
	\item Is detrainment  indeed negligible for dry thermals? \label{Q_det}
\end{enumerate}

To answer these questions we perform fully-resolved direct numerical simulation (DNS)  of three-dimensional dry thermals rising through a neutral environment, closely analogous to the laboratory experiments of S57. \note{Thermals have certainly been simulated before,  using LES \citep[e.g.][]{morrison2018,morrison2017,moser2017,langhans2015a,yeo2013,romps2010b,craig2008,stiller2001,grabowski1995,grabowski1993, grabowski1993a,grabowski1993b,grabowski1991,ogura1962, lilly1962}, but none of these works have studied the dry case in three-dimensions, and more importantly questions 1-3 above remain open.} Here we take advantage of the fact that DNS simulations have a well-defined Reynolds number, and vary it to address question \ref{Q_Re} above. We also combine our simulations with a thermal tracking algorithm inspired by that of \cite{romps2015}, which allows us to precisely define the thermal's control volume as a function of time.  Differentiating this volume with respect to height then yields the thermal's \note{(net)} entrainment rate, allowing us to address question \ref{Q_ent}. Question \ref{Q_det} will be addressed using our thermal tracking, passive tracers, as well as a density anomaly budget for the environment, which is straightforward to calculate from simulation output.

We begin by  briefly reviewing in section \ref{subsec:theory} the analytic similarity theory of S57 for dry thermals in a neutral environment.  Then in section~\ref{sec:setup} we describe our simulation setup and the broad properties of the simulated thermals. In section~\ref{sec:volume} we describe and apply a thermal tracking algorithm  to find the volume of each thermal as a function of time.  In section~\ref{sec:entrainment} we differentiate this function to find the  entrainment, verifying the  $1/r$ scaling in \eqnref{ent1}, and also verifying that detrainment in dry thermals is negligible. In section~\ref{sec:efficiency} we calculate the entrainment efficiency $e$ in our simulations, exploring its \Re\ dependence and comparing to previous work. We conclude  in section~\ref{sec:conclusions}.

%=======%
% Theory   %
%=======%

\subsection{Analytic Theory of Thermals}\label{subsec:theory}

\note{In this subsection we review the classic similarity theory of S57. This theory does not invoke the \newnote{relation}  \eqnref{turner1}, but instead assumes a geometric similarity between states of the thermal at different times, leading to the similarity constraints ~\eqnref{scorer_z} and \eqnref{scorer_w} below. These similarity constraints contain dimensionless constants which must be determined empirically. Though these constants can themselves be constrained by invoking \eqnref{turner1} or the vertical momentum equation  \citep[e.g.][]{escudier1973, turner1964b}, we nonetheless use the S57 theory to describe our thermals, for the sake of connecting with previous literature. We discuss possible elaborations to the S57 theory in the conclusions.}
% add text to conclusions setting up Brett's paper!

The S57 theory describes a  `self-similar'  thermal whose shape at any given time must be geometrically similar to its shape at any other time.\note{\footnote{This self-similar solution is supposed to hold until the thermal interacts with an environmental boundary, discontinuity, or additional lengthscale,  e.g., the thermal size becomes comparable to a scaleheight, or the thermal approaches the  tropopause in the atmosphere or a boundary in experiments or simulations.}} From this self-similarity condition S57 deduces that  the thermal's height $z$ must be linearly related to its radius $r$ as
\beqn
	z-z_0 = nr 
	\label{scorer_z}
\eeqn 
for constant $n$ and some suitably chosen `virtual origin' $z_0$. This equation just describes the cone traced out by the flanks of the thermal as it rises. 

S57 further argues, again by similarity, that the thermals' volume $V = m r^3$, for constant $m$. S57 also assumes that detrainment of parcel fluid into the environment is negligible, in which case the gross \newnote{fractional} entrainment rate $\epsilon$ is also equal to the \emph{net} \newnote{fractional} entrainment rate, which can be written as the fractional change in parcel volume with respect to height:
 \beqn
 	\epsilon \ = \ \frac{ d \ln V}{dz}\  \quad \quad \mbox{(no detrainment)} \ .
	\label{ent2}
\eeqn

Equation  \eqnref{ent2}  along with \eqnref{scorer_z} yields a $1/r$ entrainment law:
\beqn
	\epsilon \ = \ \frac{d \ln V}{dz}\  = \ 3 \frac{d \ln r}{dz} \ = \ \frac{3}{n}\frac{1}{r} \ .
	\label{scorer_ent}
\eeqn  
By measuring $r(z)$ from photographs, S57 infers from \eqnref{scorer_z} an average value of $n\equiv(dr/dz)^{-1}\approx4$, and hence $e=3/n \approx 0.75$.

S57 also argues that the Froude number $C^2$ is constant, and hence that 
\beqn
	w^2 = C^2Br \ ,
	\label{scorer_w}
\eeqn
 where $B$ is the thermal's average Archimedean buoyancy ($\meter/\sec^2$). Additionally, in the absence of detrainment the thermal's `mass anomaly' (i.e. integrated density anomaly, in kg)  is conserved. This is proportional to $BV$ and hence 
 \beqn
 	B = B_0r_0^3/r^3 \ ,
	\label{scorer_B}
 \eeqn
  where subscript ``0'' denotes initial values. Substituting \eqnref{scorer_B} and \eqnref{scorer_z} into \eqnref{scorer_w} and integrating with respect to time yields
  \beqn
  	z -z_0 = a\sqrt{t-t_0} 
	\label{scorer_soln}
\eeqn
for some suitably chosen $t_0$, where $a$ is a function of $n$, $C^2$, and $m$ (see S57 for details).  We will verify and utilize the thermal trajectory \eqnref{scorer_soln} in the course of constructing our thermal tracking algorithm. 

%======%
% setup  %
%======%

\section{Simulation Setup and Qualitative Description}\label{sec:setup}
\subsection{Simulation Setup}
We perform direct numerical simulations of dry thermals by solving the Boussinesq equations,
\beqanonum
\partial_{\tilde{t}} \tilde{\vec{u}} + \tilde{\rho}_0^{-1}\tilde{\vec{\nabla}} \tilde{p} - \tilde{\nu} \tilde{\nabla}^2\tilde{\vec{u}} + \tilde{g} \frac{\tilde{\rho}'}{\tilde{\rho}_0} \vec{e}_z &=& -\tilde{\vec{u}}\vec{\cdot}\tilde{\vec{\nabla}}\tilde{\vec{u}}, \\
\partial_{\tilde{t}} \tilde{T}' - \tilde{\kappa} \tilde{\nabla}^2 \tilde{T}' &=& -\tilde{\vec{u}}\vec{\cdot}\tilde{\vec{\nabla}} \tilde{T}', \\
\tilde{\vec{\nabla}}\vec{\cdot}\tilde{\vec{u}} &=& 0.
\eeqanonum
All dimensional variables are indicated with a tilde ($\,\tilde{\,}\,$).  Below we will non-dimensionalize the equations.  We have that $\tilde{\vec{u}}$ is the fluid velocity which is assumed to be nearly incompressible, $\tilde{p}$ is the pressure, $\tilde{g}$ is the gravitational acceleration, and $\vec{e}_z$ is the unit vector in the vertical direction (parallel to gravity). We expand the density $\tilde{\rho}$ as $\tilde{\rho}=\tilde{\rho}_0+\tilde{\rho}'$ with $\tilde{\rho}_0$ constant, and assume $\tilde{\rho}'\ll\tilde{\rho}_0$ such that density fluctuations are only important in the buoyancy term (the Boussinesq approximation).  The density changes due to temperature fluctuations $\tilde{T}'$ according to $\tilde{\rho}' = -\tilde{\alpha} \tilde{T}'$, where $\tilde{\alpha}$ is the (constant) coefficient of thermal expansion.  The viscous and thermal diffusivity are $\tilde{\nu}$ and $\tilde{\kappa}$, respectively.

The thermal is initialized as a spherical temperature perturbation with diameter $\tilde{L}_{\rm th}$ and temperature $\Delta \tilde{T}$ (but with zero velocity).  With these, we can define \textsl{dimensionless} variables, which do not have a hat,
\beqnonum
\begin{tabular}{ll}
$\tilde{\nabla}\rightarrow \left(\tilde{L}_{\rm th}^{-1}\right) \, \nabla$, &
$\tilde{T}' \rightarrow \left(\Delta \tilde{T}\right) \, T'$, \\
$\tilde{\rho}' \rightarrow \left(\tilde{\alpha} \Delta \tilde{T}\right) \, \rho'$, &
$\tilde{\vec{u}} \rightarrow \left(\tilde{u}_{\rm th}\right) \, \vec{u}$, \\
$\tilde{p} \rightarrow \left(\tilde{\rho}_0 \tilde{u}_{\rm th}^2 \right) \, p$, &
$\partial_{\tilde{t}} \rightarrow \left(\tilde{u}_{\rm th}/\tilde{L}_{\rm th}\right) \, \partial_t$,
\end{tabular}
\eeqnonum
where \note{(consistent with the energetics argument of equation~\ref{scorer_w}),}
\Beq
\tilde{u}^2_{\rm th} \equiv \frac{\tilde{g}\tilde{L}_{\rm th}\tilde{\alpha}\Delta \tilde{T}}{\tilde{\rho}_0}. \n
\Eeq
With this non-dimensionalization, the equations become
\begin{subequations}
	\Beq 
		\partial_t \vec{u} \note{+} \vec{\nabla} p - {\rm Re}^{-1} \nabla^2 \vec{u} + \rho' \vec{e}_z &=& -  \vec{u}\vec{\cdot}\vec{\nabla}\vec{u}, \label{eqn:non-dim velocity} \\
		\partial_t \rho' - {\rm Pr}^{-1}{\rm Re}^{-1} \nabla^2 \rho' &=& -\vec{u}\vec{\cdot}\vec{\nabla} \rho', \\
		\vec{\nabla}\vec{\cdot}\vec{u} &=& 0. \label{eqn:non-dim divergence}
	\Eeq
	\label{eqn:dedalus}
\end{subequations}
Now the problem is entirely characterized by only two dimensionless numbers, the Reynolds number and the Prandtl number,
\beqanonum
{\rm Re} &=& \frac{\tilde{u}_{\rm th}\tilde{L}_{\rm th}}{\nu}, \\
{\rm Pr} &=& \frac{\tilde{\nu}}{\tilde{\kappa}},
\eeqanonum
as well as the choice of noise we add to the temperature field.  In this paper, we fix ${\rm Pr}=1$, close to the atmospheric value of ${\rm Pr}=0.7$.

One of our main goals is to determine how the thermal evolution varies with ${\rm Re}$ (question \ref{Q_Re} above). We run a series of simulations with ${\rm Re} = (2/\sqrt{10})\times 10^{4} \approx 6 \, 300$, as well as a series of simulations with ${\rm Re} = (2/\sqrt{10})\times 10^3 \approx 630$.  \note{We refer to the higher Reynolds number simulations as \textit{turbulent}, as they exhibit a range of spatial scales, chaos, etc. These simulations} are intended to be representative of the turbulent entrainment found in atmospheric convection (although their Reynolds number is still much smaller than in the atmosphere). \note{In contrast, the lower Reynolds number simulations, which we term \textit{laminar}, are not chaotic and only have a single spatial scale; these simulations} may be more similar to the thermals described in simulations of ensembles of convective clouds such as \cite{hernandez2016,romps2015,sherwood2013} in which each thermal is only resolved by a handful of grid cells.\footnote{\newnote{Although our thermals could entrain via diffusion, one can check this is less efficient than advective entrainment by a factor of $\approx {\rm Re}$, and thus negligible for our simulations.}} \newnote{We also ran several simulations with ${\rm Re}\approx 1\,800$, but found them difficult to interpret as they are on the verge of turbulence, and thus exhibit large fluctuations. Thus, we will not discuss them further.}

\note{Reproducibility is important for science \citep{Oishi2018}. To aid readers to reproduce our results or to perform their own analysis of our data, we have created a github repository containing all the simulation scripts used to generate, analysis, and plot the data used in this paper. Although we could not include all the raw data from the simulations (which constitutes around 10TB of data), we do include the reduced data necessary to reproduce the plots in this paper. The repository is \url{https://github.com/lecoanet/thermals_entrainment}.}

We solve equations~\eqnref{eqn:dedalus} using the open-source Dedalus\footnote{More information at \url{dedalus-project.org}.} pseudo-spectral code \citep{burns2016}. We discretize the problem by expanding each quantity in a certain number of sine or cosine modes in $x$, $y$, and $z$.  The domain extends from $-5\tilde{L}_{\rm th}$ to  $5\tilde{L}_{\rm th}$ in the $x$ and $y$ directions, and from $0$ to $20\tilde{L}_{\rm th}$ in the $z$ direction.  For each direction, the normal velocity is expanded as a sine series, and the perpendicular velocities are expanded as cosine series.  The pressure is expanded in cosine series in all directions and the temperature perturbation is expanded in cosine series in the horizontal directions and a sine series in the vertical direction.  This corresponds to stress-free boundaries, and no-flux boundaries in the horizontal direction and isothermal boundaries in the vertical direction.  For laminar simulations, we use 256 modes in the horizontal directions and 512 modes in the vertical direction.  For turbulent simulations, we use 512 modes in the horizontal directions and 1024 modes in the vertical direction.  We use the $3/2$ padding rule when evaluating nonlinear terms to prevent aliasing errors.  To timestep the problem, we use a third order, four stage implicit-explicit Runge-Kutta timestepper \citep[][where linear terms are treated implicitly, and nonlinear terms are treated explicitly]{ascher1997}, with the timestep size set by a Courant--Friedrichs--Lewy condition with prefactor 0.7.  We run the simulations for $\approx 60$ time units, which is enough time for the thermals to approach the top boundary.

To construct the initial condition, we first specify a spherical density perturbation,
\Beq
\rho'_{\rm sph} =\frac{1}{2}\left[{\rm erf}\left(\frac{r_0 - r_{\rm IC}}{\Delta r} \right) -1\right], \n
\Eeq
where
\Beq
r_{\rm IC} \equiv \sqrt{(x-x_0)^2+(y-y_0)^2+(z-z_0)^2}, \n
\Eeq
and ${\rm erf}$ is the error function \citep[e.g.,][]{abramowitz1968}, $x_0=y_0=0$, $z_0=1.5$, and $r_0=0.5$ is the initial radius.  We take $\Delta r=0.1$ as a smoothing length. \newnote{To break the symmetries of the problem we use the initial density perturbation} $\rho'_{\rm sph}\times (1+N(x,y,z))$.  $N$ is a noise field specified in terms of the amplitude and phase of its sine and cosine modes. For each mode $\vec{k}=(k_x,k_y,k_z)$ with $k_x,k_y,k_z<128\times2\pi/10$, we set $N_{\vec{k}}$ by $A(1+k^2)^{-1/6}\xi \sin(\phi)$, where $A$ is an amplitude, $\xi$ is a normally distributed random variable and $\phi$ is a random variable uniformly distributed from $0$ to $2\pi$.  We pick $A$ such that the root-mean-square density perturbations are $0.21$.\footnote{\newnote{We also found similar results for larger $A$.}} At both Reynolds numbers, we run five simulations with different choices of the initial random seed, and thus different noise fields $N$.  This allows us to compute an ensemble average over our simulations.

As the thermal evolves, we expect its radius $r$ and typical velocity $u_{\rm rms}$ to evolve such that $r u_{\rm rms}$ stays approximately constant (since $r\sim \sqrt{t}$ and $u\sim w\sim 1/\sqrt{t}$; section~\ref{subsec:theory}).  This means that the Reynolds number of the thermal is approximately constant over the simulation.  At early times, the thermal takes up a small part of the domain, and is thus harder to resolve.  The resolution of our turbulent simulations is insufficient to fully resolve the flow at the early stages of the simulations, leading to low-amplitude Gibbs' ringing, especially in the density field. \note{In these early stages, the viscous dissipation scale is about twice the grid spacing.}  We thus limit our analysis to later stages of the simulation (e.g., $t>10$), where the thermal is larger and well-resolved \note{(i.e., the viscous dissipation scale is larger than the grid spacing).}

\subsection{Qualitative Analysis}
\label{subsec:qual_analysis}

\begin{figure*}
  \centerline{\includegraphics[width=7.1in]{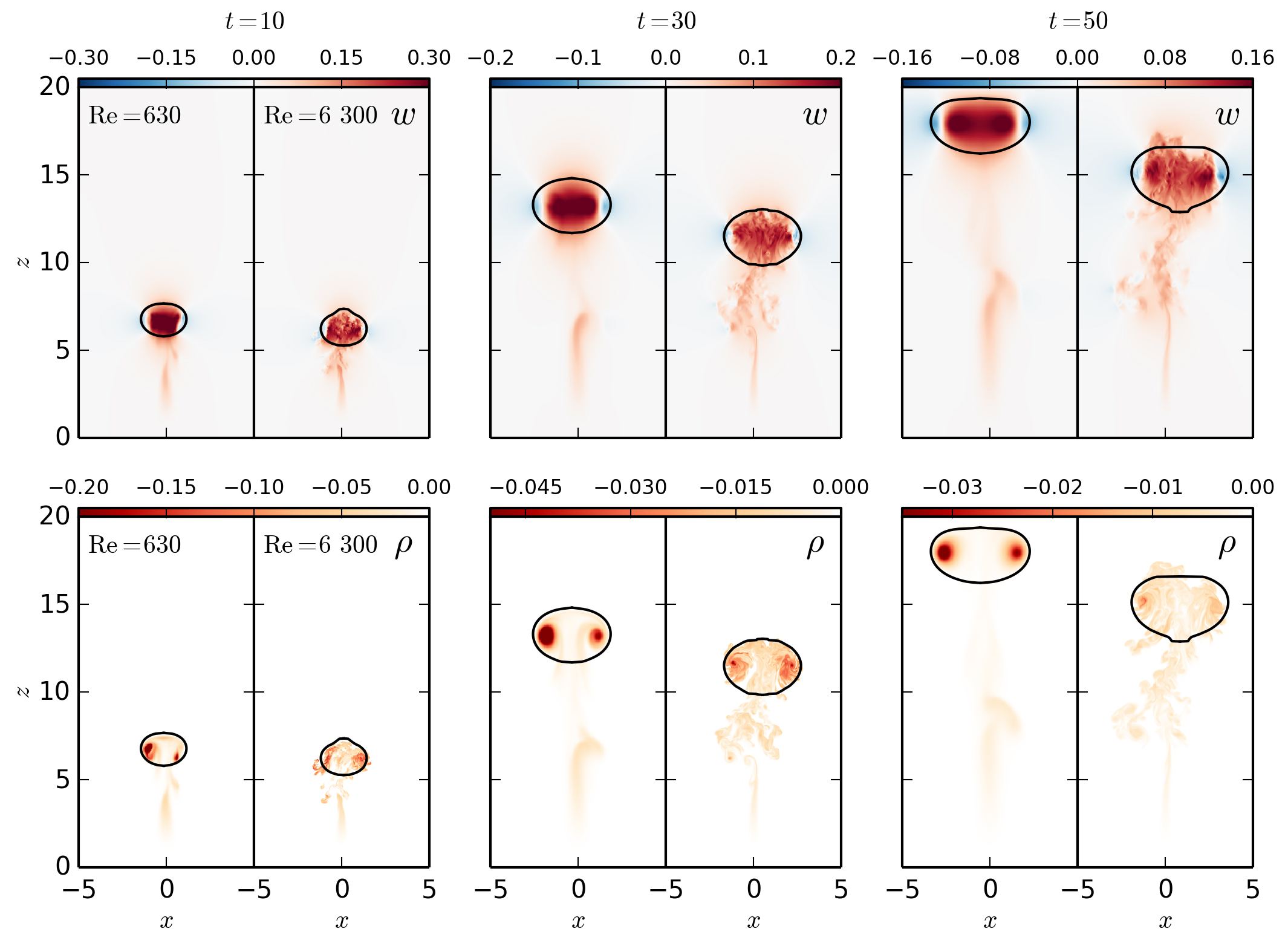}}
  \caption{2D vertical slices at $y=0$ of the vertical velocity (top) and density (bottom) of two thermals with different Reynolds numbers. The two thermals are initialized with the same noise field. At each time, the panel on the left shows the low Reynolds number thermal (${\rm Re}=630$), and the panel on the right shows the high Reynolds number thermal (${\rm Re}=6\,300$). We use the same color scale for the two thermals at the same time; we change the color scale at different times as the thermal becomes dilute and slows down over its evolution. We also plot the boundary of the thermal as computed in section~\ref{sec:volume}. The high and low \Re\ thermals exhibit comparable entrainment, with the high \Re\ case entraining slightly more. \note{The supplementary materials include movies of these two thermal simulations.}}
\label{fig:thermal_evolution}
\end{figure*}

\begin{figure}[ht]
  \centerline{\includegraphics[width=3.4in]{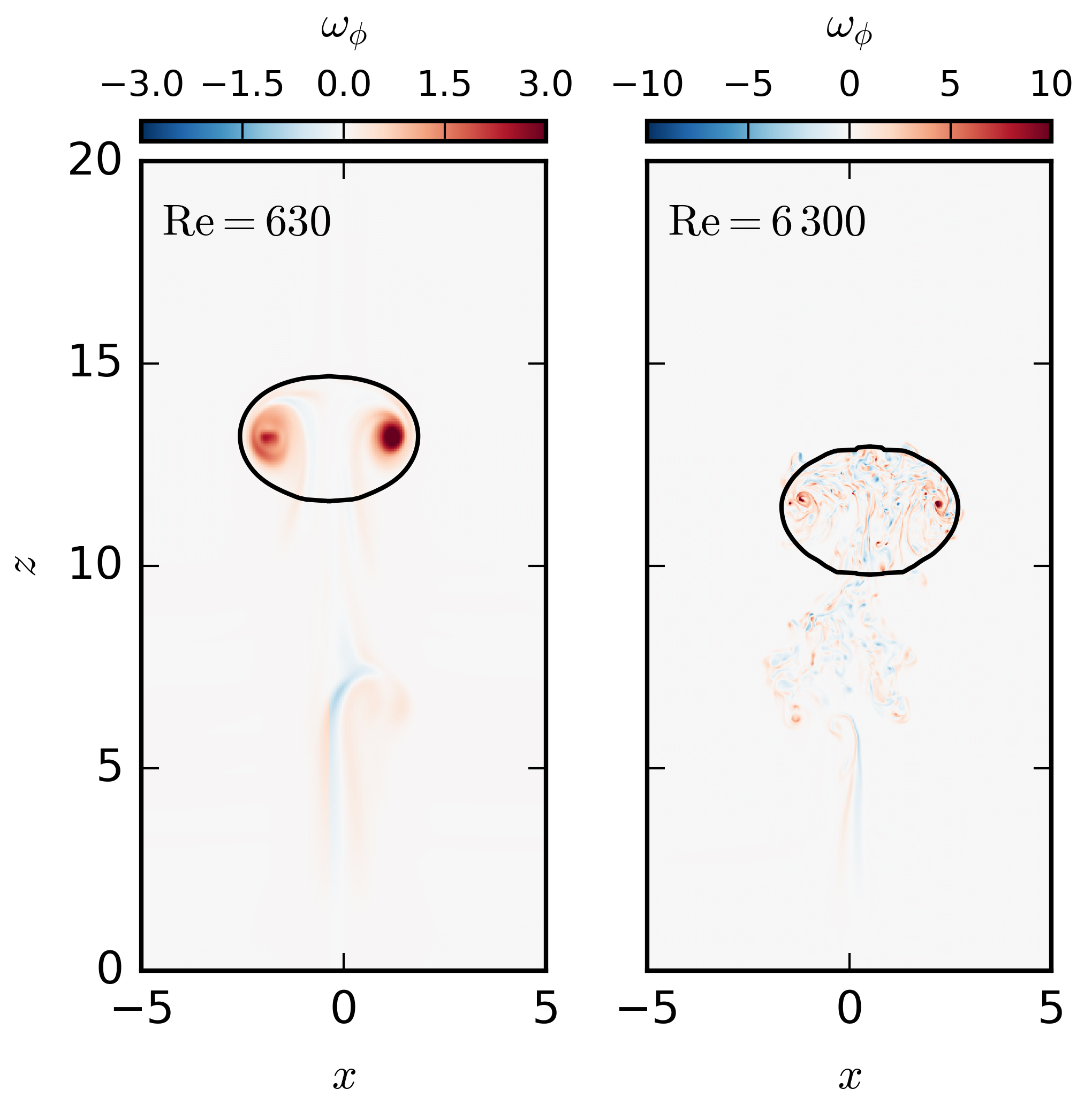}}
  \caption{2D vertical slices of $\omega_\phi$ at the $y$ thermal midpoint (defined in section~\ref{sec:volume}) of two thermals with different Reynolds numbers, at $t=30$. These slices show positive vorticity cores of the thermals, which coincide with the buoyancy anomaly. \newnote{Note the much smaller scales of $\omega_\phi$ in the turbulent case.}}
\label{fig:omega}
\end{figure}

In figure~\ref{fig:thermal_evolution} we show 2D vertical slices of the time evolution of a laminar, \Re=630  thermal  and a turbulent, \Re=6\,300 thermal (left and right panel of each pair, respectively), drawn from our ensemble of simulations and initialized with the same initial noise field. (Appendix~\ref{sec:ensemble} \newnote{discusses the} effect of changing the initial noise field on the thermal.) \note{Movies showing the evolution of these thermals are included in the supplementary information.}  Both thermals take the form of buoyant vortex rings (see figure~\ref{fig:omega}) and entrain ambient fluid as they rise, at seemingly comparable rates; this is consistent with  Turner's claim that Eqn. \eqnref{turner2} rather than turbulence drives entrainment, and provides a preliminary answer to question \ref{Q_Re} above. The thermals grow significantly larger with time, and because of the associated dilution exhibit a decrease in  vertical velocity  as well as density perturbation, as evident in the changing color scales. \note{This decrease in vertical velocity despite increasing impulse is due to `mixing drag', i.e. the fact that newly entrained, neutrally buoyant mass must accelerate by absorbing momentum from the existing thermal. (This growth and deceleration of the thermal with time contrasts the behavior of moist thermals in more realistic simulations, which we discuss this further in the conclusions.)} Both thermals also detrain, as there is a trail of fluid left below each thermal, but it is unclear from this figure how significant this is.

\note{In figure~\ref{fig:omega} we plot the azimuthal vorticity \newnote{($\omega_\phi=\partial_z u_r - \partial_r u_z$, where $\vec{u}$ is represented in cylindrical coordinates)} of these two thermals at $t=30$. Although the turbulent simulation has substantial vorticity at small scales, both simulations show strong vorticity maxima in the core of the vortex ring. Unlike, say, Hill's spherical vortex which has only positive vorticity throughout, these thermals also have some negative vorticity; thus the vertical velocity is not maximal at the center of the thermal.}

While our laminar and turbulent thermals  have much in common, differences do exist between the two cases.  The laminar thermal forms a well-defined (though somewhat asymmetric) vortex ring, whereas in the turbulent case the vortex ring structure \note{is narrower in vorticity, but more diffuse in density}. Furthermore, the turbulent thermal clearly entrains slightly more than the laminar one, leading to lower heights, vertical velocities, and smaller density anomalies  at a given time. The rest of this paper will focus on measuring this entrainment, as well as  detrainment, to yield more quantitative answers to questions \ref{Q_Re}--\ref{Q_det} above.

%=======%
% Volume  %
%=======%

\section{Thermal Volume}
\label{sec:volume}

In Figure~\ref{fig:thermal_evolution} we also plot the boundary of the thermal, \newnote{whose definition and calculation we now describe}. This boundary will be important for distinguishing the thermal from ambient fluid, as well as from its trail of detrained fluid, and will thus be critical for measuring both entrainment and detrainment.% We now briefly describe the algorithm for identifying this boundary; full details are given in  Appendix~\ref{sec:tracking}.

\newnote{Heuristically, we think of a thermal as a region of fluid that moves coherently. Following \citet{romps2015}, we define the thermal to be the axisymmetric volume whose averaged vertical velocity matches the velocity of the of the thermal's top.\footnote{It is relatively easy to identify the top of the thermal, as there is a sharp change in density at the top of the thermal.} This definition agrees with the well-known notion  of the vortex ring `bubble' or `atmosphere' \citep[][sec 3.2]{akhmetov2009,shariff1992}. This definition is based on fundamental fluid dynamics and treats thermals as coherent dynamic entities, and as such is arguably preferable (at least in this context) to conventional definitions of cloud control volumes based on arbitrary thresholds of cloud condensate, vertical velocity, and/or buoyancy \citep[e.g.][]{derooy2013,dawe2011, romps2010b}. Similar tracking algorithms to ours have been used in the recent studies of \cite{hernandez2018,morrison2018,hernandez2016,sherwood2013}. Our algorithm seems to work well for our simulations (see movies in the supplementary information), but is of course not perfect, and sometimes misclassifies fluid as being outside the thermal, even though it continues to rise coherently with the rest of the thermal volume. This complicates our detrainment calculation in section~\ref{sec:entrainment}.}

%Our thermal tracking algorithm is inspired by \cite{romps2015}, \note{which defines the thermal volume boundary as the zero contour of the azimuthally symmetric Stokes streamfunction, defined in Eqn. \eqnref{eqn:streamfunction} below. This definition agrees with the well-known notion  of the vortex ring `bubble' or `atmosphere' \citep[][sec 3.2]{akhmetov2009,shariff1992}, and delineates the region of fluid that moves coherently with the vortex ring, and whose volume-averaged velocity equals the gross translational velocity of the ring itself. This definition is based on fundamental fluid dynamics and treats thermals as coherent dynamic entities, and as such is arguably preferable (at least in this context) to conventional definitions of cloud control volumes based on arbitrary thresholds of cloud condensate, vertical velocity, and/or buoyancy \citep[e.g.][]{derooy2013,dawe2011, romps2010b}. Similar tracking algorithms to ours have been used in the recent studies of \cite{hernandez2018,morrison2018,hernandez2016,sherwood2013}. Our algorithm seems to work well for our simulations (see movies in the supplementary information), but is of course not perfect, and sometimes misclassifies fluid as being outside the thermal, even though it continues to rise coherently with the rest of the thermal volume. This complicates our detrainment calculation in section~\ref{sec:entrainment}.}

\newnote{We now briefly describe the algorithm for identifying this boundary; full details are given in  Appendix~\ref{sec:tracking}. The first step is to calculate the position of the top of the thermal, $z_{\rm t}$, which we do using a density threshold. To calculate the thermal-top velocity $w_{\rm top}$, we first perform a fit to $z_{\rm t}$, and then take the derivative analytically.}

\begin{figure}
  \centerline{\includegraphics{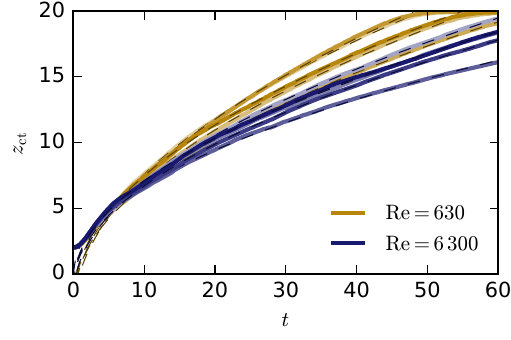}}
  \caption{\newnote{Thermal}-top height as a function of time in our ten thermal simulations. Lower Reynolds number are shown in yellow, and higher Reynolds number are shown in blue. With dashed lines, we also plot the $\sqrt{t}$ fit given in equation~\ref{eqn:z_ct fit}. The fit is very good for times greater than $\approx 10$.}
\label{fig:z_ct}
\end{figure}

\begin{figure*}
  \centerline{\includegraphics[width=7.1in]{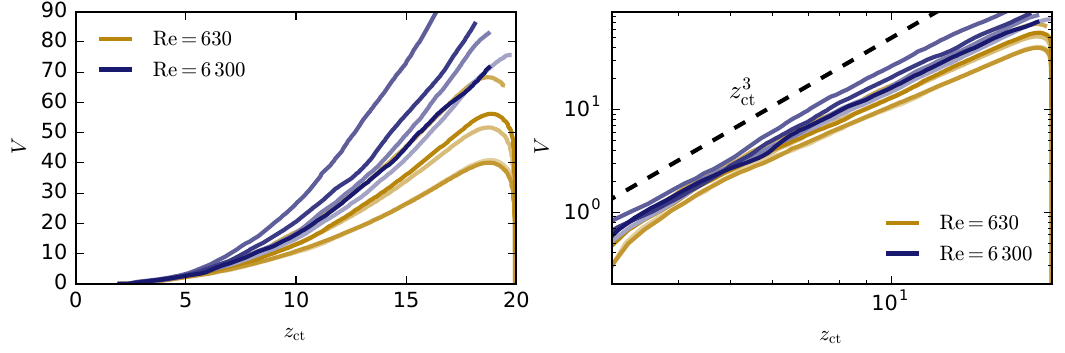}}
  \caption{Thermal volume as a function of \newnote{thermal}-top height in our ten thermal simulations. Lower Reynolds number are shown in yellow, and higher Reynolds number are shown in blue. The left panel uses a linear scale; the right panel uses a log scale to illustrate that the volume increases like $z_{\rm t}^3$.  The decrease in volume at late times in our lower Reynolds number simulations are due to the thermal interacting with the top boundary.}
\label{fig:volume}
\end{figure*}

In figure~\ref{fig:z_ct}, we plot the \newnote{thermal}-top height as a function of time for each of our simulations.  The curves have negative concavity, again indicating deceleration.  The turbulent thermals have lower heights than the laminar thermals at the same time, consistent with figures~\ref{fig:thermal_evolution} \& \ref{fig:ensemble}.  For each curve, we also plot the fit \newnote{we use to calculate $w_{\rm top}$} (equation~\eqref{eqn:z_ct fit}).  The fit is very good except at early times $t<10$.  This is expected, as it takes some time for the thermal to spin up and reach the self-similar regime.  Furthermore, as described earlier, our higher Reynolds number simulations are not very well-resolved for $t<10$---although this is not expected to lead to large differences in bulk properties like the \newnote{thermal}-top height.  The good fit to the $\sqrt{t}$ time-dependence gives a first quantitative check of the classical theory of S57 (section~\ref{subsec:theory}).  Because the thermal boundary is calculated based on $z_{\rm t}$ and our fit is not good for $t<10$, we expect the thermal properties will not be accurately calculated during these early stages of the simulations.

\newnote{To find the axisymmetric volume translating with this velocity, we calculate the Stokes stream-function $\psi$ associated with the the azimuthally averaged velocities, $\vec{u}_{\rm axi}=(u_{\rm axi}(r,z),w_{\rm axi}(r,z))$. The Stokes stream-function satisfies}
\Beq
\label{eqn:streamfunction}
\vec{\nabla}\vec{\times}\left(\psi \vec{e}_\phi\right) = (2\pi r)(\vec{u}_{\rm axi} - w_{\rm top} \vec{e}_z),
\Eeq
where $\vec{e}_\phi$ is the unit vector in the azimuthal direction.  This is the stream-function associated with the thermal velocity, in the frame moving upwards at the \newnote{thermal}-top velocity.  The thermal boundary is the line of constant $\psi$ starting from the point $(r,z)=(0,z_{\rm top})$ where the axisymmetric vertical velocity matches the \newnote{thermal}-top velocity, i.e., $w_{\rm axi}(0,z_{\rm top})=w_{\rm top}$.  Thus, by definition, the thermal has no net mass flux in the frame moving with the \newnote{thermal}-top velocity, and its average vertical velocity in the rest frame is equal to the \newnote{thermal}-top velocity. More details about this calculation can be found in Appendix~\ref{sec:tracking}.

Figure~\ref{fig:thermal_evolution} shows our calculation of the thermal boundary in our simulations.  This automated procedure seems to do a good job of identifying the thermal, and seems to match the region one might pick out ``by eye.'' When visualizing the simulation data, it is easiest to identify the thermal when looking at the vertical velocity (rather than the density), because it is smoother.  The vertical velocities are strong and coherent within the thermal volume, as opposed to the density field which is highly variable and exhibits pockets of completely unmixed ambient fluid within the thermal (see e.g. the $t=30$ panel of the Re=6300 simulation in figure~\ref{fig:thermal_evolution}). This makes it more difficult to determine the thermal volume by only looking at the density field.  Note that  these pockets of unmixed environmental fluid seem to result from engulfment \note{at the rear of the thermal} by the large-scale  vortex ring  circulation, and are only later mixed/diffused in, as also noted by previous authors  \citep{moser2017,johari1992, sanchez1989, turner1964, woodward1959, scorer1957}. This is consistent with entrainment  originating from Eqn. \eqnref{turner2}, rather than  turbulence.

The boundary of the thermal  tells us the radius of the thermal as a function of height, $r(z)$.  We can then calculate the thermal volume by simply summing $\pi r(z)^2 dz$ over $z$. This thermal volume is shown in figure~\ref{fig:volume}.  The classical S57 theory predicts that $z\sim r$ and hence that volume should increase like $V \sim z_{\rm t}^{3}$, which we also find in our simulations.  The turbulent thermals have larger volumes than the laminar thermals at the same height,  again suggesting somewhat larger entrainment rates for the turbulent thermals. There is also a nonintuitive \emph{decrease} in the volume in our lower Reynolds number simulations   when $z_{\rm t}$ becomes large, but this is an artifact of the thermal interacting with the top boundary (see figure~\ref{fig:z_ct}). 

%Since the goal of this work is to study thermals far away from boundaries (as is the case in the atmosphere),  we do not try to track the thermals after they reach the top boundary.

%===========%
% entrainment   %
%===========%

\section{Net Entrainment and detrainment}\label{sec:entrainment}
While the increasing volume of our thermals with time must be  due to entrainment of environmental fluid, it can also in principle be offset by  detrainment of fluid from the thermal to the environment. This possibility is not so far-fetched, as we clearly see in the simulations a trail of detrained fluid below the thermal (figure~\ref{fig:thermal_evolution}). Accordingly, and analogously to $\epsilon$, we define the fractional \emph{detrainment rate} $\delta$ to be the fraction of the thermal's volume that it detrains per unit vertical distance traveled, again in units of $\meter\inverse$. We may then generalize \eqnref{ent2} by allowing for detrainment and properly equating  $d \ln V/dz$ to  the  \emph{net} fractional entrainment rate \enet, giving
\Beq
\enet \ \equiv\  \frac{d\ln V}{dz} \  = \  \epsilon - \delta \ .
\Eeq
 Since \enet\ can be calculated from the volume only, we focus on it first; estimating $\delta$ will require additional machinery, to be discussed later. 
 
One goal of our paper is to check the $\enet \sim 1/r$ scaling suggested by  Eqn. \eqnref{scorer_ent}.  To do so,  we must calculate $r$ and $d \ln V/dz$  in our simulations.  We define the thermal radius $r_{\rm th}$ to be the maximum radius of the thermal boundary.  To calculate the entrainment rate, we calculate $dV/dt$ with a second-order central differencing formula (as implemented in \verb!numpy!'s \verb!gradient! function).  Then we calculate \enet\  by dividing $dV/dt$ by $V w_{\rm top}$. Taking the derivative via finite differences introduces noise in the entrainment rate. Although we could have reduced this noise by fitting the data and taking a derivative analytically, or by smoothing, we decided to minimize the data processing to eliminate any biases so introduced. On the other hand, it seems acceptable to fit the \newnote{thermal}-top heights via \eqnref{eqn:z_ct fit} because it is in an early stage of the data analysis.

\begin{figure}
  \centerline{\includegraphics[width=3.4in]{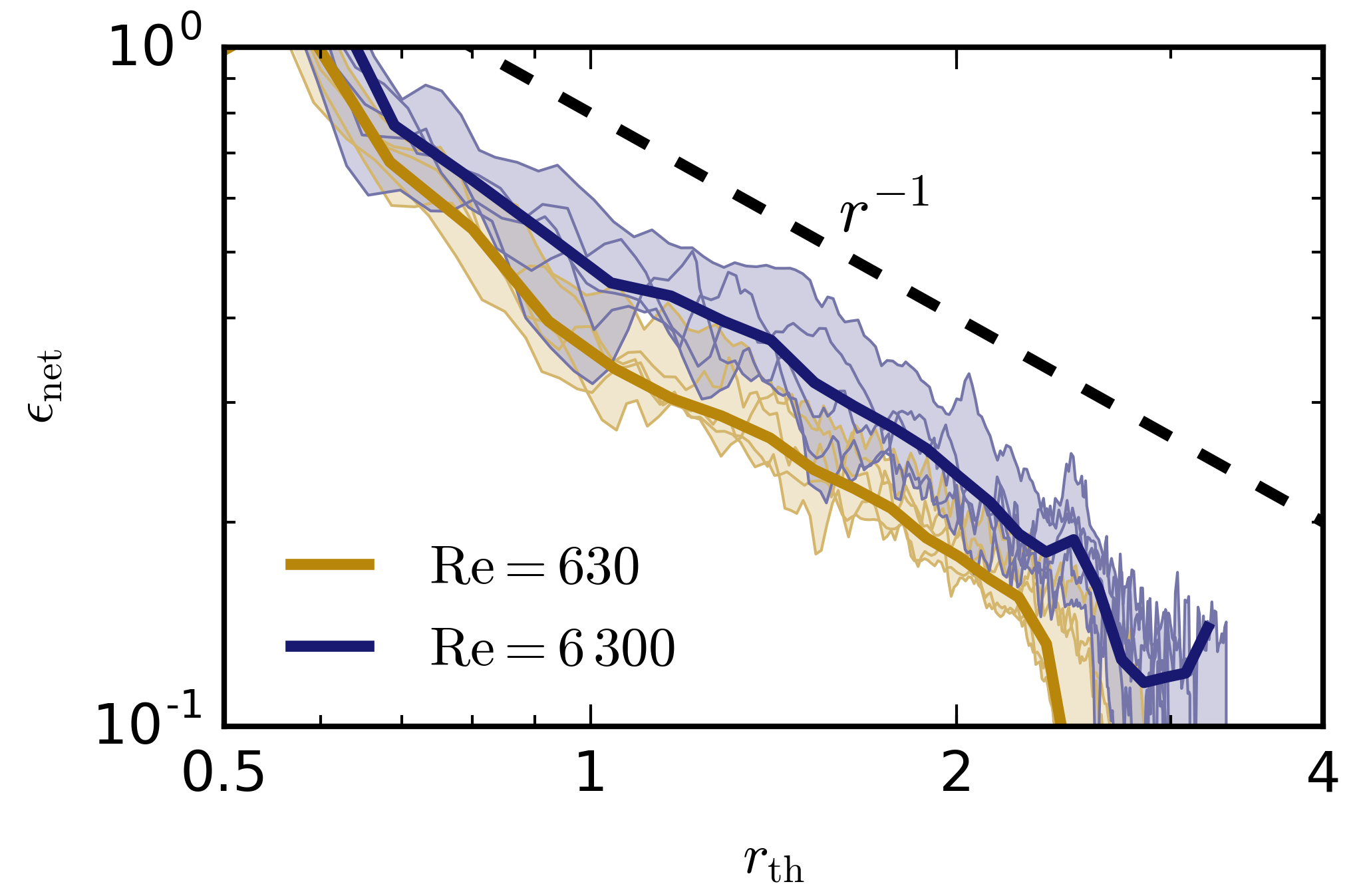}}
  \caption{Net entrainment rate as a function of radius.  The mean net entrainment rate across the thermals with a given Reynolds number is plotted in the dark thick line, and each individual simulation is plotted in a thin line.  The range between simulations of a given Reynolds number is shaded. The entrainment rate decreases like $r^{-1}$ over the course of the simulation.}
\label{fig:entrainment}
\end{figure}

Figure~\ref{fig:entrainment} shows the net entrainment rate \enet\ as a function of thermal radius \rth\ on log-log scales. A slope of -1 is evident,  providing a direct verification of the $1/r$ scaling of \enet.\footnote{\note{Although the thermal is still spinning up for $r\lesssim1$, the $1/r$ scaling appears to roughly hold even then.}}  We find this dependence for both individual simulations (thin lines), and the mean (thick lines; calculated by binning \rth\   in steps of $0.1$, and then calculating the average \enet\  and average $\rth$ over simulations with a given Reynolds number in each bin). Note that the higher Reynolds number thermals indeed have slightly larger net entrainment rates than the lower Reynolds number thermals. 
There is significantly more scatter in entrainment rates among the turbulent thermals than the laminar thermals, likely due to turbulent chaos.

Next we turn to detrainment. \note{A simple approach to measuring detrainment would be to diagnose changes in the thermal's mass anomaly $BV$  with time. This approach, however, assumes that $B$ is uniform throughout the thermal, so that a change in $BV$ can be confidently associated with a change in $V$. Such an assumption seems reasonable for our turbulent thermals, but not for the laminar ones (Fig. \ref{fig:thermal_evolution}).  Also, if some mass anomaly leaves our idealized, axisymmetric  thermal volume at some time but  then re-enters it soon afterwards and continues to move upwards, this would still be measured as detrainment, which seems undesirable.}

\note{To deal with these complications, we take the following approach. In the turbulent case we integrate $\rho'$ in the  \emph{environment below the thermal} as a measure of detrained mass anomaly $M_d$, and track the rate at which $M_d$ changes. We only track detrained mass anomaly below the thermal because the truly detrained fluid rises less quickly than the thermal, and thus falls below it. This method for measuring detrainment neglects the  significant mass anomaly exiting the thermal volume's sides and top which subsequently re-enters the thermal from the sides (in contrast to entrained fluid, which enters from below). We think of this fluid as always contained within some ``platonic'' thermal, and never as having really been detrained. The fact that it leaves our axisymmetric thermal volume is an imperfection of our idealized definition of the thermal boundary. As for the laminar case, because $B$ is not uniform in this case we felt compelled to use a different method entirely. For this case we use a passive tracer initialized within the thermals at several times within each simulation, and track the rate at which the tracer is detrained into the environment. Full details of these calculations are given in Appendix \ref{sec:detrainment}. }

\begin{figure}
  \centerline{\includegraphics{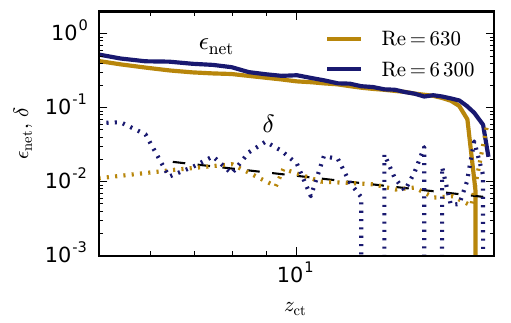}}
  \caption{Average net entrainment and detrainment rates as a function of the height of the thermal. The solid lines show the net entrainment rate, and the dotted lines show the detrainment rate.  Because the detrainment rate is typically at least an  order of magnitude smaller than the net entrainment rate, we have $\epsilon_{\rm net}\approx \epsilon$. The dashed black line shows a $z^{-1}$ power law.}
\label{fig:detrainment}
\end{figure}

With these methodologies in place, we  calculate detrainment rates $\delta$ and directly compare to net entrainment rates \enet.  Ensemble averages of these quantities are shown in figure~\ref{fig:detrainment} (the averages are taken in the same way as for figure~\ref{fig:entrainment}, but with \zct\ bins of size 0.5). Both the laminar and turbulent detrainment rates are an order of magnitude smaller than the entrainment rates,  justifying the approximation of S57 and others in neglecting detrainment. Furthermore, the detrainment rates, despite being calculated using different algorithms, are similar for the two sets of calculations, and thus also only marginally sensitive to \Re. However, it does seem that the detrainment process itself depends on \Re\ --- it is steady and primarily through the bottom for the laminar thermals, but intermittent and primarily through the sides for the turbulent thermals. \note{In contrast, in both cases the entrainment is predominantly from the bottom of the thermal.}  Note also that the detrainment rate for the laminar thermal decreases like $z^{-1}$, and since $z_{\rm t}\sim r_{\rm th}$, this suggests that $\delta\sim 1/r$, just like the net entrainment rate. The detrainment rate for the turbulent thermals is too noisy to determine a temporal variation, but we hypothesize that it may also decrease like $r^{-1}$.  Figures \ref{fig:entrainment} and \ref{fig:detrainment}  thus show that $\enet \sim 1/r$ in all cases and $\delta \sim 1/r$ in the laminar case, which in conjunction with $\epsilon = \enet + \delta$  provides strong support for the entrainment assumption \eqnref{ent1}.

%==========%
% efficiency    %
%==========%

\section{Entrainment Efficiency} \label{sec:efficiency}
\begin{figure}
  \centerline{\includegraphics[width=3.4in]{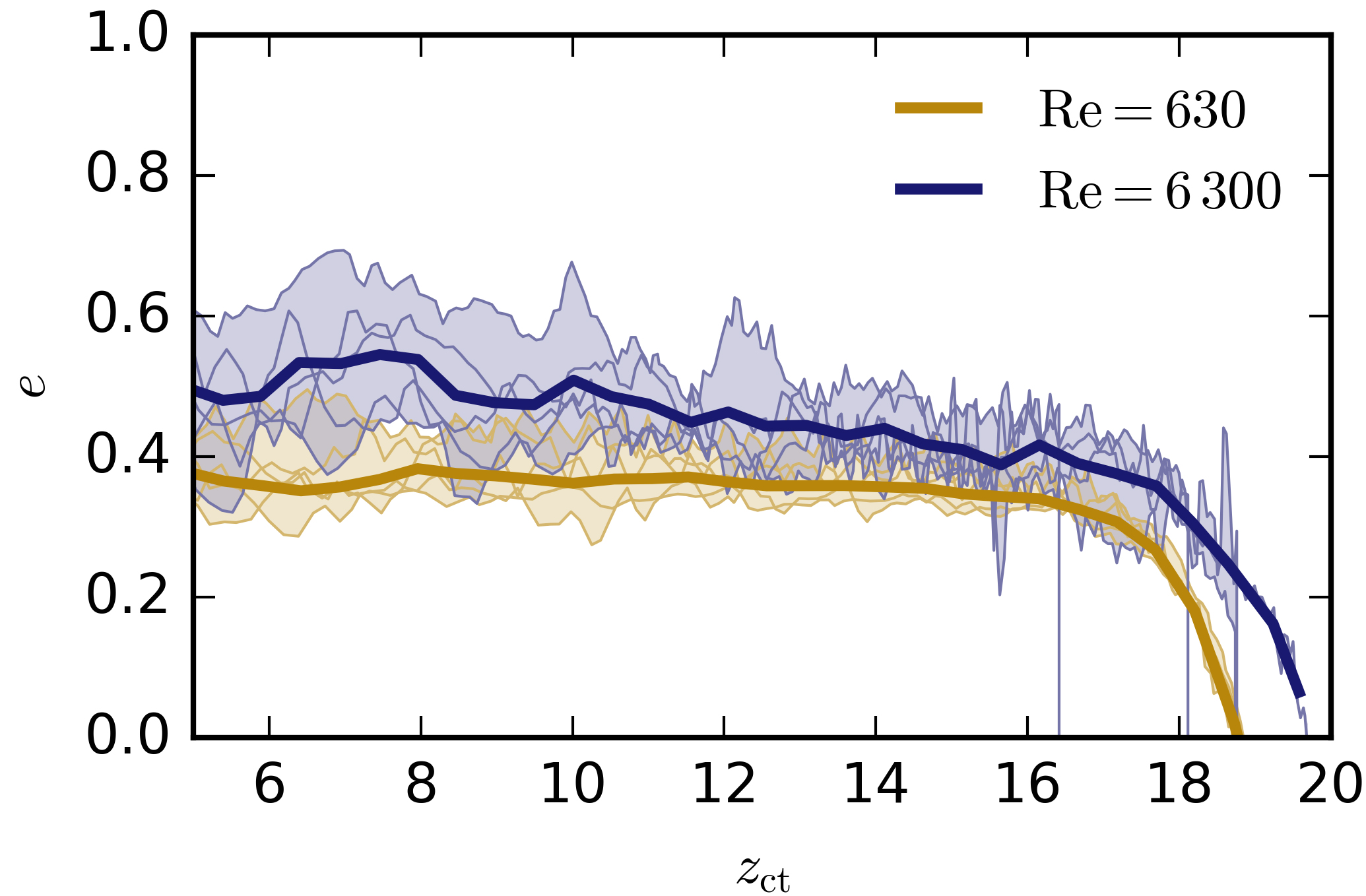}}
  \caption{Entrainment efficiency $e$ (equation~\ref{ent1}) as a function of \newnote{thermal}-top height.  The mean efficiency across the thermals with a given Reynolds number is plotted in the dark thick line, and each individual simulation is plotted in a thin line.  The range between simulations of a given Reynolds number is shaded. Although there is a wide spread in the entrainment efficiency in any individual simulation (especially for higher Reynolds number), the turbulent thermals typically have higher entrainment efficiency than the laminar thermals.}
\label{fig:efficiency}
\end{figure}

Since figure \ref{fig:detrainment} shows that $\delta \ll \enet$ and hence that $\epsilon \approx \enet$, we may estimate the entrainment efficiency $e$ in \eqnref{ent1} by multiplying  $\enet$ by $r_{\rm th}$.  This should be approximately constant over the course of a simulation.  We plot $e$ as a function of time in figure~\ref{fig:efficiency}. We find that the averaged entrainment efficiency (dark thick lines) is indeed roughly constant with time, except at the very beginning of the simulations or near the end when the thermals begin to interact with the top boundary.  Furthermore, the turbulent thermals on average have a larger entrainment efficiency than the laminar thermals.  However, there is substantial spread between the different simulations (especially for the turbulent thermals), and an individual turbulent thermal may have a smaller entrainment efficiency at a given time than an individual laminar thermal.

The average entrainment efficiency drops quickly for heights $z_{\rm t}>16$ for our laminar simulations \note{and for heights $z_{\rm t}>18$ for our turbulent simulations}.  As described above, this is because the thermals are approaching the top boundary of the simulation domain.  However, there is a slower, more secular decrease in the entrainment efficiency with height, for both Reynolds numbers.  \newnote{Daniel: did we want to edit/excise this next sentence ?} For the turbulent thermals, this may be partially explained by a single thermal whose entrainment efficiency appears to be decreasing with time more robustly than the other thermals \note{in the ensemble}. This is not explained by the  S57  theory (section~\ref{subsec:theory}).  Nevertheless, it does seem that the lowest order description of the problem is that the entrainment efficiency is roughly constant.

To more quantitatively compare the entrainment efficiencies of the simulations with different Reynolds numbers, we also take a time average of the entrainment efficiency between $z_{\rm t}=6$ and $z_{\rm t}=16$.  The ensures we are not influenced by initial transient effects at early stages of the simulations, nor by interactions with the boundaries at late stages.  The values are reported in table~\ref{tab:entrainment}. We find that the laminar thermals have a lower entrainment efficiency than the turbulent thermals by only about 20\%, yielding a quantitative answer to question \ref{Q_Re}. \note{The entrainment efficiencies of our turbulent simulations are similar to those found in laboratory experiments with an initial aspect ratio $\geq 1$, but smaller than those of R61, S57, and MTT56 which had an initial aspect ratio of 0.5. This is quantitatively consistent with the dependence on aspect ratio found in  \cite{Lai2015}.}  Also note that entrainment efficiencies of order slightly less than 1 are broadly consistent with typical fractional entrainment rates of $O( (1\ \km)\inverse )$ measured in simulations and observations \citep[e.g.][]{derooy2013,malkus1954}, since a typical radius for clouds is $r  \approx O(1 \km)$.

\begin{table}
\caption{The average entrainment efficiency ($e$; see equation~\eqnref{ent1}) in our simulations at two Reynolds numbers, and as reported in laboratory experiments.  We also provide the equivalent $n$ parameter described in S57 (see section~\ref{subsec:theory}), \note{as well as the initial aspect ratio}.}
\label{tab:entrainment}
\centering

%\begin{tabular}{cccl}
%\toprule
%Thermals & $e$ & $n$ & $\delta$ \\
%\midrule
%${\rm Re} = 630$ & $0.36$ & $8.3$ & $0.02$ \\
%${\rm Re} = 6\,300$ & $0.47$ & $6.4$ & $0.003$ \\
%R61, S57, MTT56 & $\sim 0.75$ & $\sim 4$ & $0$ \\
%\bottomrule
%\end{tabular}
%\end{table}

\begin{tabular}{ccrc}
\toprule
Thermals & $e$ & $n$ & $L/H$ \\
\midrule
${\rm Re} = 630$ & $0.36$ & $8.3$ & 1  \\
${\rm Re} = 6\,300$ & $0.47$ & $6.4$  & 1 \\
R61, S57, MTT56 & $\sim 0.75$ & $\sim 4$ & 0.5 \\
\citet{Bond2010} & $\sim 0.5$ & $\sim 6$ & 2 \\
\bottomrule
\end{tabular}
\end{table}

%===========%
% Conclusions  %
%===========%

\section{Discussion \& Conclusions}\label{sec:conclusions}

In this paper we present a suite of direct numerical simulations of resolved, dry thermals. We initialize the thermals with a sphere of buoyant fluid released at the bottom of our simulation domain, with random perturbations to break symmetries. We run an ensemble of simulations with different random perturbations (but with the same statistical properties), and with  Reynolds numbers of either $\Re\approx 630$ (laminar) or $\Re\approx 6\,300$ (turbulent). To quantitatively measure the entrainment and detrainment associated with each thermal, we implement a thermal tracking algorithm inspired by \cite{romps2015}, where the resulting thermal boundary is shown with a black line in figure~\ref{fig:thermal_evolution}. 

Our entrainment calculation allows us to directly verify the long-standing entrainment assumption $\epsilon=e/r$ for both laminar and turbulent thermals. Furthermore, we find that both cases entrain comparably, with entrainment efficiencies $e$ that differ by only 20\%. This is consistent with the idea of entrainment governed by \eqnref{turner2} for buoyant vortex rings \citep{turner1957}.  We also found that detrainment rate is roughly an order of magnitude smaller than the net entrainment rate, suggesting that \emph{net} entrainment rates are approximately equal to \emph{gross} entrainment rates.

%One main goal of this work was to compare laminar and turbulent thermals.  Even our laminar thermals are hundreds of grid points across, a resolution that cannot be afforded in more complicated large-eddy simulations (LES) of cloud ensembles.  Thermals in such LES  likely entrain/detrain primarily via sub-grid turbulent diffusion schemes or via numerical diffusion, rather than resolved turbulence.  Thus, they may more closely resemble our laminar thermals rather than our turbulent thermals. 

The finding that entrainment has a strong laminar component should be contrasted with the plume picture of convection, in which entrainment is thought of as purely turbulent \citep[e.g.][]{turner1986,squires1962,kuo1962,morton1956}. Future work will investigate whether the 20\% increase in $e$ in the turbulent case stems from turbulence in an essential way, or if turbulence merely changes the large-scale characteristics of the flow (such as the circulation $K$ in \eqnref{turner2}) which could also change $e$. Also note that entrainment governed by \eqnref{turner2} is distinct from the `dynamic' entrainment proposed by \cite{houghton1951} and incorporated in later studies \citep[e.g.][]{morrison2017,derooy2010,ferrier1989,asai1967}, since dynamic entrainment is a consequence of \emph{positive} vertical acceleration, whereas we see negative vertical acceleration here.

\note{While the results presented here help paint a picture of entrainment in dry convection, a key question is how these results extend to atmospheric moist convection. As mentioned in Section \ref{subsec:qual_analysis}, simulated moist atmospheric thermals \note{(without resolved turbulent mixing)} exhibit a fairly constant size and vertical velocity \citep[][]{hernandez2016,romps2015,sherwood2013}, in contrast to our dry thermals. These moist thermals also appear to have a qualitatively different momentum budget than dry thermals, with little to no mixing drag  but a significant perturbation pressure drag \citep{morrison2018, romps2015, sherwood2013, deroode2012}; this contrasts with the dry thermal's momentum budget, which can be completely characterized by mixing drag and virtual mass effects \citep{turner1964b}.}

What might be the sources of these differences? Regarding entrainment/detrainment, the fact that moist thermals exhibit a constant size implies that $\enet \approx 0$ and hence $\delta \approx \epsilon$. Since our gross entrainment rates seem to be similar to those measured in more realistic simulations, the discrepancy is likely due to detrainment. There are two likely candidates for a sharp increase in detrainment in more realistic cases. One is the stable stratification of the real atmosphere, which reduces the buoyancy of unsaturated parcels as they rise. The other candidate is the mixing of cloudy and clear air, which causes evaporation of condensate and again a decrease in buoyancy. These processes ultimately generate negatively buoyant air which is likely to detrain, a process formalized in the `buoyancy-sorting' scheme of \cite{raymond1986}. 

\note{As for dynamics, the argument of \cite{turner1957} relies on a constant circulation $K$, which itself relies on the existence  of a circuit through the center of the vortex ring along which the density anomaly is negligible. Moist thermals, however, can exhibit self-generated buoyancy in their center due to latent heat release \citep[e.g.][]{morrison2018,romps2015}, so such a circuit may not exist. To the contrary, that buoyancy in the thermal's center may very well \emph{increase} the circulation $K$, which by \eqnref{turner1} would help $r$ to increase more slowly or even stay constant in the face of increasing $P$, as is indeed observed. Future work could assess how the constraint \eqnref{turner1} operates in the moist case, and also assess 
how a stable stratification and condensate evaporation contribute to  buoyancy sorting and enhanced detrainment. }

\note{Another direction for future work would be to develop a theory for thermals based not on the similarity assumptions \eqnref{scorer_z} and \eqnref{scorer_w}, but rather on the \newnote{relation} \eqnref{turner1} \emph{in addition to} the vertical momentum equation. \cite{turner1957} employed the former but not the latter, and \cite{turner1964b} and \cite{escudier1973} employed the latter but not the former. Thus,  it is still possible that using both might result in a complete theory for dry thermals with no additional, empirical constants.}

\section*{Acknowledgments}

DL is supported by a  PCTS fellowship and a Lyman Spitzer Jr. fellowship. NJ was supported by the Visiting Scientist Program in Princeton's Atmosphere and Ocean Science program, as well as a Harry Hess fellowship from the Princeton Geoscience Department. Computations were conducted with support by the NASA High End Computing (HEC) Program through the NASA Advanced Supercomputing (NAS) Division at Ames Research Center on Pleiades with allocations GID s1647 and s1439.

%=========%
% Appendix   %
%=========%

\appendix

%=========%
% Ensemble  %
%=========%
\section{Ensemble Characteristics}\label{sec:ensemble}

\begin{figure*}
  \centerline{\includegraphics[width=7.1in]{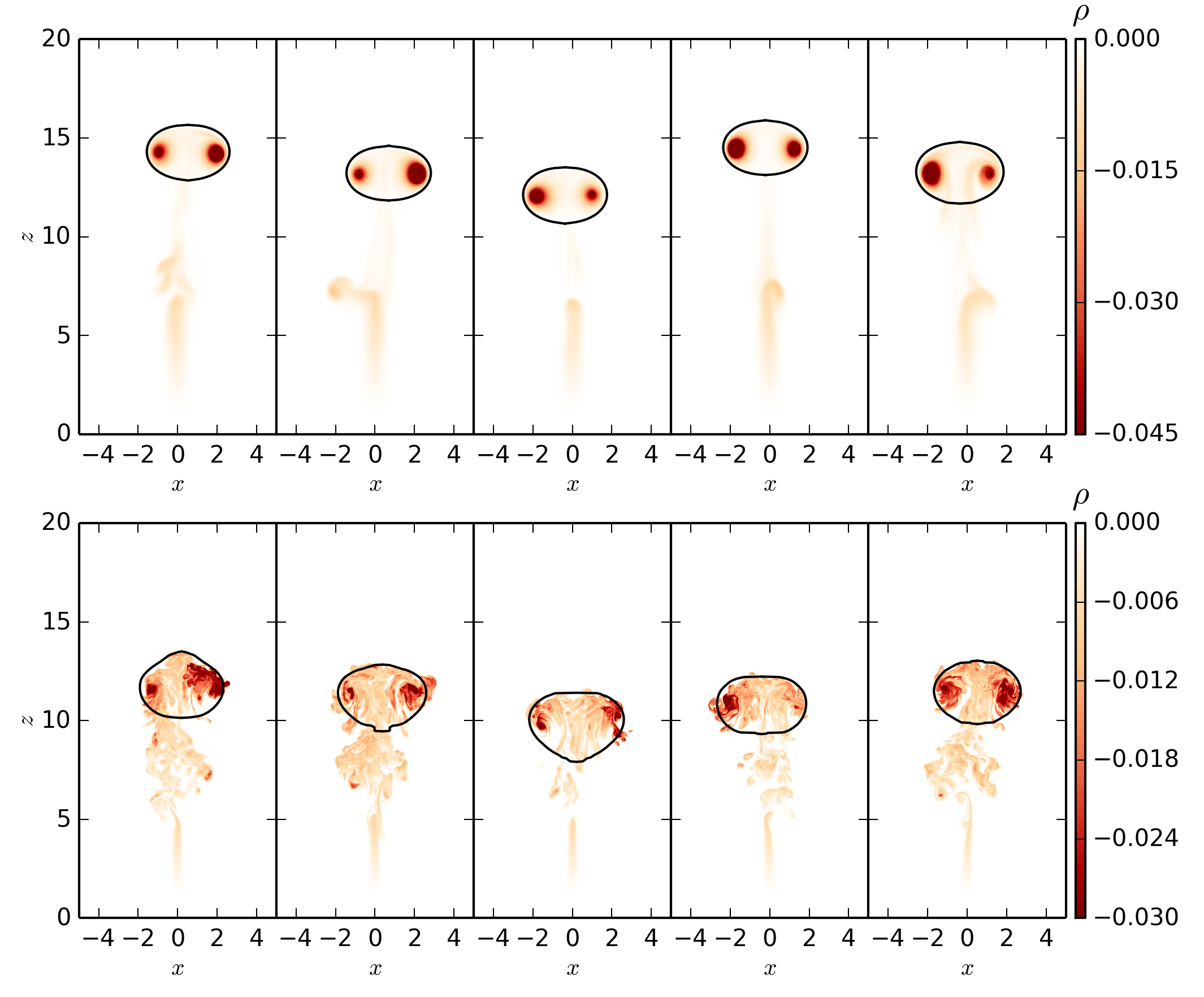}}
  \caption{2D vertical slices at $y=0$ of density perturbation of thermals at $t=30$ for low Reynolds number (top) and high Reynolds number (bottom). Each vertical pair of simulations has identical initial noise fields, but this noise field varies in the simulations arranged horizontally, leading to substantial variation in thermal evolution across simulations at a given Reynolds number. However, the laminar thermals are systematically higher and have larger density anomalies at this time than the turbulent thermals. We also plot the boundary of the thermal as computed in section~\ref{sec:volume}.}
\label{fig:ensemble}
\end{figure*}

We find that different initial noise fields can lead to different thermal evolution, even though the statistical properties of the noise (e.g., power spectrum, root-mean-square fluctuations) are identical.  Figure~\ref{fig:ensemble} shows 2D vertical slices of the density in all ten of our simulations at $t=30$.  Each vertical pair represents simulations with the same initial condition; the top plots show laminar thermals and the bottom plots show turbulent thermals. The rightmost plots are from the same simulations as shown in figure~\ref{fig:thermal_evolution}.

Although there is substantial variation between the different simulations with the same Reynolds number, we can still clearly identify some trends with the Reynolds number.  In particular, as we described for the single choice of initial noise field in figure~\ref{fig:thermal_evolution}, the laminar thermals systematically rise higher and have larger density anomalies than the turbulent thermals.  Thus this observation is robust to the exact choice of initial noise.

%=========%
% Tracking    %
%=========%

\section{Thermal Tracking Algorithm}\label{sec:tracking}

Here we will describe the details of our thermal tracking algorithm. The first step is to determine the \newnote{thermal}-top height $z_{\rm t}(t)$. At each time (we have full volume outputs every $1/\sqrt{10}\approx 0.36$ time units), we first calculate the horizontal average of $\rho'$ at each height, $\langle \rho'\rangle_{x,y}$. We define a cutoff value of $0.1\max_z \langle \rho'\rangle_{x,y}$. \note{The height is very insensitive to the cutoff value because the density field is very sharp.} The \note{thermal}-top height is the highest point at which $\langle\rho'\rangle_{x,y}$ is greater than this cutoff value in absolute value.

\newnote{We find that the volume of the thermal shows sensitivity to the value of $w_{\rm top}$, so care must be taken when calculating $dz_{\rm t}/dt$. We find the best way to calculate this derivative is to first fit
\Beq\label{eqn:z_ct fit}
z_{\rm t} \approx a \sqrt{t} + z_0,
\Eeq
based on \eqnref{scorer_soln}, using a non-linear least squares fit, and then take the derivative analytically to determine $w_{\rm top}$.}

Next we calculate the horizontal midpoint of the thermal.  We define the midpoint using the vertical velocity.  At each height, we define $x_m$ by
\Beq
x_m = \frac{\sum_{w>0} x w}{\sum_{w>0} w},
\Eeq
and similarly for $y_m$.  We then pick the vertical height which maximizes $\sum_{w>0} w$. The horizontal midpoint of the thermal is then the $(x_m,y_m)$ at this height. \note{Using a mass-weighting yields a similar midpoint.}

We then calculate the azimuthal average of the vertical velocity at every vertical height. To do this, we define a radial grid which has the same grid spacing as the $x$ or $y$ grids, but only extends from $r=0$ to $r=5$. To calculate $w_{\rm axi}(r,z)$, we take the average of $w$ in azimuthal rings of width $\Delta r$ centered around $r+\Delta r/2$.

Once we have $w_{\rm axi}(r,z)$, we can calculate the Stokes stream-function $\psi$ satisfying Eqn. \eqnref{eqn:streamfunction}.  We do this using Dedalus \citep{burns2016}.  First we (spectrally) interpolate $w_{\rm axi}(r,z)$ onto a Chebyshev grid in the radial direction.  Then we solve the linear boundary value problem,
\Beq
\partial_r\psi = 2\pi r \left(w_{\rm axi} -w_{\rm top}\right),
\Eeq
with the boundary condition $\psi(r=0)=0$.

The boundary of the thermal is then the contour $\psi=0$.  To find this contour, at every height, we find the maximum of $\psi$.  If the maximum occurs at radius larger than $0.18$ and is positive, then we use root finding to find a zero of $\psi$ at larger radii than the maximum.  This defines the radius of the thermal at each height.  The lower cutoff of $0.18$ is used to \note{ensure the thermal is one simply connected volume, and} prevent the thermal tracking from thinking that pockets of detrained fluid near the midpoint are part of the thermal.  In one simulation, there were several times at which such pockets of fluid were identified as being part of the thermal (even though they were far below and disconnected from the real thermal). In this case, we manually removed these spurious thermal components (i.e., set the radius of the thermal at these heights to zero).

%===========%
% detrainment   %
%===========%

\section{Detrainment}\label{sec:detrainment}
Here we detail our detrainment calculations. These are somewhat involved, as the detrainment rate is low, which makes it difficult to calculate. Furthermore  the physical process differs between laminar and turbulent thermals, leading us to different methodologies for the two cases.

\note{We conceptually think of a thermal as a coherent volume of fluid that is rising at similar velocities. Then for fluid to be detrained, it must be rising slower than the thermal, and thus} must ultimately end up below the thermal. In the turbulent case, however, we find that often fluid initially leaves the thermal volume on the tops and sides, and that some of this fluid is subsequently reincorporated into the thermal. Although one might think of this process as a detrainment followed by subsequent entrainment event, we instead think of this fluid as being part of the thermal the whole time, even if it was not correctly identified as such by our thermal tracking algorithm. This is a limitation of our assumption of axisymmetry when defining the thermal volume.

Thus, to diagnose detrainment in turbulent thermals, we note that the tail of detrained fluid (e.g., as seen in figure~\ref{fig:thermal_evolution}) has similar density to the average density of the thermal (this does not seem to be the case for laminar thermals). If we assume the average density of the detrained fluid matches the average density of the thermal, then the volume of detrained fluid can be measured by the mass anomaly of fluid detrained (integral of $\rho'$ over that volume). Since the background $\rho$ is not even defined in our simulations due to the non-dimensionalization, we will refer in this appendix to such mass anomalies simply as `mass'.

Because the environmental fluid has $\rho'=0$ (from the Boussinesq approximation), entrainment of environmental fluid cannot change the mass of the thermal $M_{\rm th}\equiv \rhoth V$, \note{where $\rhoth$ is the average density perturbation of the thermal}. Entrainment changes $\rhoth$ and $V$ \note{individually} but not their product. Thus, if we know the total mass of detrained fluid \Md\ throughout the simulation,  we may approximate  the detrainment rate as 
\Beq\label{eqn:detrainment}
\delta \approx  \frac{1}{M_{\rm th}} \frac{d M_{\rm d}}{dz} \ .
\Eeq
 
Defining \Md\ requires care, however. If we simply integrate $\rho'$ over the whole domain excluding the thermal, we integrate not only over detrained fluid below the thermal but  also fluid above or on the sides as well (e.g., above the thermal at $t=50$), much of which will subsequently be re-entrained. With this in mind, we define the detrained mass \Md\ to be only the mass  below $z_{\rm bot}$, the lowest point in the thermal volume.  Once a detrained fluid element is below the thermal, it will very rarely be re-entrained \note{because it is rising less quickly than the thermal}. (Recall that in our Boussinesq model, the background has $\rho'=0$, so does not contribute to \Md.) We also calculated the mass below $z_{\rm bot}-r_{\rm th}/4$. Although there is less mass below that point than below $z_{\rm bot}$ at any given time, it leads to similar detrainment rates. This suggests that almost all of the mass that falls below $z_{\rm bot}$ will soon fall below $z_{\rm bot}-r_{\rm th}/4$, i.e., that it has truly been detrained.

\begin{figure}
  \centerline{\includegraphics[width=3.4in]{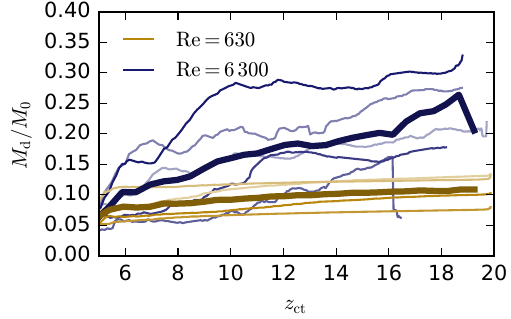}}
  \caption{The fraction of detrained mass, defined as mass below $z_{\rm bot}$, the lowest point of the thermal, as a function of \newnote{thermal}-top height. The blue curves show turbulent thermals, and the yellow curves show laminar thermals. In dark thick lines, we show the detrained mass averaged over our ensemble of turbulent or laminar thermals.}
\label{fig:mass_below}
\end{figure}

Figure~\ref{fig:mass_below} shows in thin lines \Md\  for all of our simulations, normalized by the total initial mass, $M_0=-(4\pi/3)r_0^3$. In the turbulent simulations, the signal is noisy; there can be abrupt increases in the detrained mass, as detrainment events are often discrete, and the detrained mass occasionally decreases as the thermal rises, due to some of the mass that falls below $z_{\rm bot}$  becoming re-entrained. To alleviate some of these issues, we also plot the ensemble-average detrained mass(dark thick lines in figure~\ref{fig:mass_below}, computed as in figure~\ref{fig:entrainment} but with \zct\ bins of size 0.5).

In its early evolution ($z_{\rm t}<5$, not shown), the thermal detrains a non-negligible  amount of fluid, roughly 5-10\% of $M_0$.  After this initial transient, however, there is on average a relatively slow growth of detrained mass.  Indeed, on average the detrained mass only grows by a factor of 4--5  from $\zct=5$ through the end of the simulation for turbulent thermals. In contrast, the volume changes by almost a factor of $100$ (figure~\ref{fig:volume}), suggesting detrainment is indeed much weaker than entrainment.

Figure~\ref{fig:mass_below} also shows the detrained mass for laminar thermals, which increases continuously and smoothly as the thermal rises. The laminar thermals detrain much less mass than the turbulent thermals. However, the detrained fluid in laminar thermals (see the fluid immediately below the thermal in figure~\ref{fig:thermal_evolution}) is much less dense than the average density of the thermal. This is because laminar thermals' density is much more concentrated in a vortex core than turbulent thermals' density. \note{Thus we cannot approximate the detrained \textit{volume} to be proportional to the detrained mass for laminar thermals.}

To measure the detrainment rate for laminar thermals, then, we instead run a set of dye simulations which track the movement of a passive scalar (``dye'') both into and out of the thermal. We solve the equation
\Beq\label{eqn:dye}
\partial_t s - D \nabla^2 s = -\vec{u}\vec{\cdot}\vec{\nabla} s
\Eeq
for the passive scalar field $s$. We use Schmidt number of unity ($D=\nu$). $s$ is represented as a cosine series in the $x$ and $y$ directions, and a sine series in the $z$ direction. \note{The field $s$ denotes the density of the passive tracer; we will use $c$ to denote a volume integral of (i.e., amount of) the passive tracer.}

Every $\sqrt{10}$ time units, we reinitialize $s$ to be \note{nearly} unity within the thermal, and \note{nearly} zero outside, \note{as described below}. To define the thermal, we use the thermal radius as a function of $z$ calculated with our volume tracker (section~\ref{sec:volume}). We use this to define the radius of the thermal as a function of angle from the vertical from the center of the thermal---the vertical center of the thermal is defined as the average of the highest and lowest points in the thermal. Then for each grid point in the domain, we calculate the angle to the thermal center relative to the vertical, and then calculate the radius of the thermal for that angle $r_{\rm th}(\theta)$ using linear interpolation. Then the dye field is initialized to
\Beq\label{eqn:dye_IC}
s_{\rm init}(r,\theta) = \frac{1}{2}\left[1-{\rm erf}\left(\frac{r-r_{\rm th}(\theta)}{\Delta r}\right)\right],
\Eeq
where $\Delta r=0.1$ is a smoothing length.

After each reinitialization, the dye field is advected along with the thermal. Almost all of the dye stays within the thermal, but is diluted due to the entrainment of environmental fluid. If we calculate the entrainment rate by measuring the amount of environmental fluid that enters the volume, we find entrainment consistent with our measurements purely in terms of the thermal volume. This is because entrainment is much stronger than detrainment, so the entrainment rate is almost entirely determined by the change in volume. There is a small amount of dye which is initialized within the thermal that leaves the thermal. Some dye leaves the thermal volume due to diffusion, but some is advected out, i.e., is detrained. As above, we consider dye detrained if it is below the lowest point of the thermal. We denote amount of dye below the lowest point of the thermal by $c_{\rm d}$.

\begin{figure}
  \centerline{\includegraphics[width=3.4in]{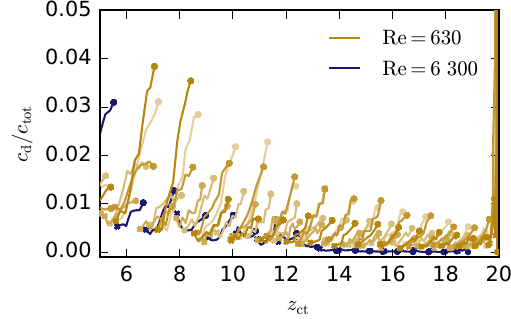}}
  \caption{The fraction of detrained dye, defined as the dye below $z_{\rm bot}$, the lowest point of the thermal as a function of \newnote{thermal}-top height. The dye field is reinitialized to be almost entirely within the thermal every $\sqrt{10}$ time units, denoted with a cross. The final detrained dye fraction before the next reinitialization is denoted with a circle. The yellow curves show the fraction for laminar thermals, the blue curve for one turbulent thermal. The slope of the line connecting the cross and circle for each curve is the detrainment rate. We are unable to measure the detrainment rate for turbulent thermals using this method because it takes more than $\sqrt{10}$ time units for the detrained fluid to fall below $z_{\rm bot}$.}
\label{fig:c_dye}
\end{figure}

To quantify detrainment, we calculate how $c_{\rm d}$ changes relative to the total amount of dye, $c_{\rm tot}$. The total dye is approximately equal to the volume of the thermal at the previous reinitialization time. We plot the fraction of detrained dye, $c_{\rm d}/c_{\rm tot}$, in figure~\ref{fig:c_dye}. The fraction of detrained dye at reinitialization is marked with a cross and the last fraction recorded before the following reinitialization is marked with a circle. There is a small amount of detrained dye immediately after reinitialization because the dye field is not set to exactly zero outside the thermal (equation~\ref{eqn:dye_IC}). Sometimes the dye near the boundaries is advected into the thermal, causing the amount of detrained dye to decrease in time. Nevertheless the amount of detrained dye mostly increases until the dye field is reinitialized.

We then define the detrainment rate for laminar thermals to be
\Beq
\delta = \frac{1}{c_{\rm tot}}\frac{\Delta c_{\rm d}}{\Delta z_{\rm t}}, \n
\Eeq
where $\Delta c_{\rm d}$ and $\Delta z_{\rm t}$ are the change in the detrained dye and height over the course of one reinitialization step (of time $\sqrt{10}$). Thus, this derivative represents the slope of the line between the cross and circle of each line segment in figure~\ref{fig:c_dye}. We use this ``course-grained'' derivative because there are transients associated with the reinitialization which makes it difficult to interpret derivatives on finer scales. Since $\delta$ is a function of $z_{\rm t}$, we interpret our calculated value as corresponding to the height $z_{\rm t, 0} + \Delta z_{\rm t}/2$.

Figure~\ref{fig:c_dye} also shows the fraction of detrained dye for a turbulent simulation (corresponding to the rightmost simulation in figure~\ref{fig:ensemble}). Strikingly, this calculation shows almost no detrainment for $z_{\rm t}>13$. This appears to contradict figure~\ref{fig:thermal_evolution}, which appears to show detrained fluid below the turbulent thermal. Turbulent thermals have very little detrained dye because it takes longer than $\sqrt{10}$ time units for dye which leaves the thermal (typically from the top or the sides) to fall below the lowest point of the thermal. The detrained fluid is then reinitialized to having $c=0$ before it can fall below the thermal. This suggests the calculation needs more ``memory'' in order to correctly calculate detrainment. However, this would require longer times between reinitialization, which decreases the time resolution of the detrainment rate. For these reasons, we appeal to the assumption that the density of detrained fluid is equal to the average density of the thermal to calculate detrainment for turbulent thermals.

%===========%
% Bibliography %
%===========%

\bibliographystyle{wileyqj}
\bibliography{library_correct}

\begin{thebibliography}{79}
\providecommand{\natexlab}[1]{#1}
\providecommand{\url}[1]{\texttt{#1}}
\providecommand{\urlprefix}{URL }
\expandafter\ifx\csname urlstyle\endcsname\relax
  \providecommand{\doi}[1]{doi:\discretionary{}{}{}#1}\else
  \providecommand{\doi}{doi:\discretionary{}{}{}\begingroup
  \urlstyle{rm}\Url}\fi

\bibitem[{{Abramowitz} and {Stegun}(1968)}]{abramowitz1968}
{Abramowitz} M, {Stegun} IA. 1968. \emph{{Handbook of mathematical functions
  with formulas, graphs and mathematical tables}}. Dover.

\bibitem[{Akhmetov(2009)}]{akhmetov2009}
Akhmetov DG. 2009. \emph{{Vortex rings}}. Springer Berlin Heidelberg: Berlin,
  Heidelberg, \doi{10.1007/978-3-642-05016-9}.

\bibitem[{Asai and Kasahara(1967)}]{asai1967}
Asai T, Kasahara A. 1967. {A Theoretical Study of the Compensating Downward
  Motions Associated with Cumulus Clouds}. \emph{Journal of the Atmospheric
  Sciences} \textbf{24}(5): 487--496,
  \doi{10.1175/1520-0469(1967)024<0487:ATSOTC>2.0.CO;2}.

\bibitem[{Ascher \emph{et~al.}(1997)Ascher, Ruuth and Spiteri}]{ascher1997}
Ascher UM, Ruuth SJ, Spiteri RJ. 1997. {Implicit-explicit Runge-Kutta methods
  for time-dependent partial differential equations}. \emph{Applied Numerical
  Mathematics} \textbf{25}(2-3): 151--167, \doi{10.1016/S0168-9274(97)00056-1}.

\bibitem[{Batchelor(2000)}]{batchelor2000}
Batchelor GK. 2000. \emph{{An Introduction to Fluid Dynamics}}. Cambridge
  University Press.

\bibitem[{Blyth \emph{et~al.}(2005)Blyth, Lasher-Trapp and Cooper}]{blyth2005}
Blyth AM, Lasher-Trapp SG, Cooper WA. 2005. {A study of thermals in cumulus
  clouds}. \emph{Quarterly Journal of the Royal Meteorological Society}
  \textbf{131}(607): 1171--1190, \doi{10.1256/qj.03.180}.

\bibitem[{Bond and Johari(2010)}]{Bond2010}
Bond D, Johari H. 2010. {Impact of buoyancy on vortex ring development in the
  near field}. \emph{Experiments in Fluids} \textbf{48}(5): 737--745,
  \doi{10.1007/s00348-009-0761-z}.

\bibitem[{Burns \emph{et~al.}(2016)Burns, Vasil, Oishi, Lecoanet and
  Brown}]{burns2016}
Burns KJ, Vasil GM, Oishi JS, Lecoanet D, Brown B. 2016. {Dedalus: Flexible
  framework for spectrally solving differential equations}. \emph{Astrophysics
  Source Code Library} .

\bibitem[{Carpenter \emph{et~al.}(1998)Carpenter, Droegemeier and
  Blyth}]{carpenter1998}
Carpenter RL, Droegemeier KK, Blyth AM. 1998. {Entrainment and Detrainment in
  Numerically Simulated Cumulus Congestus Clouds. Part II: Cloud Budgets}.
  \emph{Journal of the Atmospheric Sciences} \textbf{55}(23): 3433--3439,
  \doi{10.1175/1520-0469(1998)055<3433:EADINS>2.0.CO;2}.

\bibitem[{Cotton(1975)}]{cotton1975}
Cotton WR. 1975. {Theoretical cumulus dynamics}. \emph{Reviews of Geophysics}
  \textbf{13}(2): 419--448, \doi{10.1029/RG013i002p00419}.

\bibitem[{Craig and D{\"{o}}rnbrack(2008)}]{craig2008}
Craig GC, D{\"{o}}rnbrack A. 2008. {Entrainment in Cumulus Clouds: What
  Resolution is Cloud-Resolving?} \emph{Journal of the Atmospheric Sciences}
  \textbf{65}(12): 3978--3988, \doi{10.1175/2008JAS2613.1}.

\bibitem[{Damiani \emph{et~al.}(2006)Damiani, Vali and Haimov}]{damiani2006}
Damiani R, Vali G, Haimov S. 2006. {The Structure of Thermals in Cumulus from
  Airborne Dual-Doppler Radar Observations}. \emph{Journal of the Atmospheric
  Sciences} \textbf{63}(5): 1432--1450, \doi{10.1175/JAS3701.1}.

\bibitem[{Dawe and Austin(2010)}]{dawe2011}
Dawe JT, Austin PH. 2010. {Interpolation of LES Cloud Surfaces for Use in
  Direct Calculations of Entrainment and Detrainment}. \emph{Monthly Weather
  Review} \textbf{139}(2): 444--456, \doi{10.1175/2010mwr3473.1}.

\bibitem[{Dawe and Austin(2013)}]{dawe2013}
Dawe JT, Austin PH. 2013. {Direct entrainment and detrainment rate
  distributions of individual shallow cumulus clouds in an les}.
  \emph{Atmospheric Chemistry and Physics} \textbf{13}(15): 7795--7811,
  \doi{10.5194/acp-13-7795-2013}.

\bibitem[{de~Roode \emph{et~al.}(2012)de~Roode, Siebesma, Jonker and
  de~Voogd}]{deroode2012}
de~Roode SR, Siebesma aP, Jonker HJJ, de~Voogd Y. 2012. {Parameterization of
  the Vertical Velocity Equation for Shallow Cumulus Clouds}. \emph{Monthly
  Weather Review} \textbf{140}(8): 2424--2436, \doi{10.1175/MWR-D-11-00277.1}.

\bibitem[{de~Rooy \emph{et~al.}(2013)de~Rooy, Bechtold, Fr{\"{o}}hlich,
  Hohenegger, Jonker, Mironov, {Pier Siebesma}, Teixeira and Yano}]{derooy2013}
de~Rooy WC, Bechtold P, Fr{\"{o}}hlich K, Hohenegger C, Jonker H, Mironov D,
  {Pier Siebesma} A, Teixeira J, Yano JI. 2013. {Entrainment and detrainment in
  cumulus convection: an overview}. \emph{Quarterly Journal of the Royal
  Meteorological Society} \textbf{139}(670): 1--19, \doi{10.1002/qj.1959}.

\bibitem[{de~Rooy and Siebesma(2010)}]{derooy2010}
de~Rooy WC, Siebesma AP. 2010. {Analytical expressions for entrainment and
  detrainment in cumulus convection}. \emph{Quarterly Journal of the Royal
  Meteorological Society} \textbf{136}(650): 1216--1227, \doi{10.1002/qj.640}.

\bibitem[{Escudier and Maxworthy(1973)}]{escudier1973}
Escudier MP, Maxworthy T. 1973. {On the motion of turbulent thermals}.
  \emph{Journal of Fluid Mechanics} \textbf{61}(3): 541--552,
  \doi{10.1017/S0022112073000856}.

\bibitem[{Ferrier and Houze(1989)}]{ferrier1989}
Ferrier BS, Houze RA. 1989. {One-Dimensional Time-Dependent Modeling of GATE
  Cumulonimbus Convection}. \emph{Journal of the Atmospheric Sciences}
  \textbf{46}(3): 330--352,
  \doi{10.1175/1520-0469(1989)046<0330:ODTDMO>2.0.CO;2}.

\bibitem[{Fohl(1967)}]{fohl1967}
Fohl T. 1967. {Turbulent Effects in the Formation of Buoyant Vortex Rings}.
  \emph{Journal of Applied Physics} \textbf{38}(10): 4097--4098,
  \doi{10.1063/1.1709085}.

\bibitem[{Grabowski(1993)}]{grabowski1993}
Grabowski WW. 1993. {Cumulus entrainment, fine‐scale mixing, and buoyancy
  reversal}. \emph{Quarterly Journal of the Royal Meteorological Society}
  \textbf{119}(513): 935--956, \doi{10.1002/qj.49711951305}.

\bibitem[{Grabowski(1995)}]{grabowski1995}
Grabowski WW. 1995. {Entrainment and mixing in buoyancy‐reversing convection
  with applications to cloud‐top entrainment instability}. \emph{Quarterly
  Journal of the Royal Meteorological Society} \textbf{121}(522): 231--253,
  \doi{10.1002/qj.49712152202}.

\bibitem[{Grabowski \emph{et~al.}(1991)Grabowski, Clark, Grabowski and
  Clark}]{grabowski1991}
Grabowski WW, Clark TL, Grabowski WW, Clark TL. 1991. {Cloud–Environment
  Interface Instability: Rising Thermal Calculations in Two Spatial
  Dimensions}. \emph{Journal of the Atmospheric Sciences} \textbf{48}(4):
  527--546, \doi{10.1175/1520-0469(1991)048<0527:CIIRTC>2.0.CO;2}.

\bibitem[{Grabowski \emph{et~al.}(1993{\natexlab{a}})Grabowski, Clark,
  Grabowski and Clark}]{grabowski1993a}
Grabowski WW, Clark TL, Grabowski WW, Clark TL. 1993{\natexlab{a}}.
  {Cloud-Environment Interface Instability: Part II: Extension to Three Spatial
  Dimensions}. \emph{Journal of the Atmospheric Sciences} \textbf{50}(4):
  555--573, \doi{10.1175/1520-0469(1993)050<0555:CEIIPI>2.0.CO;2}.

\bibitem[{Grabowski \emph{et~al.}(1993{\natexlab{b}})Grabowski, Clark,
  Grabowski and Clark}]{grabowski1993b}
Grabowski WW, Clark TL, Grabowski WW, Clark TL. 1993{\natexlab{b}}.
  {Cloud-Environment Interface Instability. Part III: Direct Influence of
  Environmental Shear}. \emph{Journal of the Atmospheric Sciences}
  \textbf{50}(23): 3821--3828,
  \doi{10.1175/1520-0469(1993)050<3821:CEIIPI>2.0.CO;2}.

\bibitem[{Held(2005)}]{held2005}
Held IM. 2005. {The gap between simulation and understanding in climate
  modeling}. \emph{Bulletin of the American Meteorological Society}
  \textbf{86}(11): 1609--1614, \doi{10.1175/BAMS-86-11-1609}.

\bibitem[{Hernandez-Deckers and Sherwood(2016)}]{hernandez2016}
Hernandez-Deckers D, Sherwood SC. 2016. {A Numerical Investigation of Cumulus
  Thermals}. \emph{Journal of the Atmospheric Sciences} \textbf{73}(10):
  4117--4136, \doi{10.1175/JAS-D-15-0385.1}.

\bibitem[{Hernandez-Deckers and Sherwood(2018)}]{hernandez2018}
Hernandez-Deckers D, Sherwood SC. 2018. {On the Role of Entrainment in the Fate
  of Cumulus Thermals}. \emph{Journal of the Atmospheric Sciences}
  \textbf{75}(11): 3911--3924, \doi{10.1175/JAS-D-18-0077.1}.

\bibitem[{Heus \emph{et~al.}(2009)Heus, Jonker, {Van den Akker}, Griffith,
  Koutek and Post}]{heus2009}
Heus T, Jonker HJJ, {Van den Akker} HEA, Griffith EJ, Koutek M, Post FH. 2009.
  {A statistical approach to the life cycle analysis of cumulus clouds selected
  in a virtual reality environment}. \emph{Journal of Geophysical Research}
  \textbf{114}(D6): D06\,208, \doi{10.1029/2008JD010917}.

\bibitem[{Houghton and Cramer(1951)}]{houghton1951}
Houghton HG, Cramer HE. 1951. {A Theory of entrainment in convective currents}.
  \emph{Journal of Meteorology} \textbf{8}(2): 95--102,
  \doi{10.1175/1520-0469(1951)008<0095:ATOEIC>2.0.CO;2}.

\bibitem[{Jeevanjee \emph{et~al.}(2017)Jeevanjee, Hassanzadeh, Hill and
  Sheshadri}]{jeevanjee2017a}
Jeevanjee N, Hassanzadeh P, Hill S, Sheshadri A. 2017. {A perspective on
  climate model hierarchies}. \emph{Journal of Advances in Modeling Earth
  Systems} \textbf{9}(4): 1760--1771, \doi{10.1002/2017MS001038}.

\bibitem[{Johari(1992)}]{johari1992}
Johari H. 1992. {Mixing in Thermals with and without Buoyancy Reversal}.
  \doi{10.1175/1520-0469(1992)049<1412:MITWAW>2.0.CO;2}.

\bibitem[{Klocke \emph{et~al.}(2011)Klocke, Pincus and Quaas}]{klocke2011}
Klocke D, Pincus R, Quaas J. 2011. {On constraining estimates of climate
  sensitivity with present-day observations through model weighting}.
  \emph{Journal of Climate} \textbf{24}(23): 6092--6099,
  \doi{10.1175/2011JCLI4193.1}.

\bibitem[{Kuo(1962)}]{kuo1962}
Kuo H. 1962. {On the Controlling Influences of Eddy Diffusion on Thermal
  Convection}. \emph{Journal of Atmospheric Sciences} \textbf{19}: 236--243.

\bibitem[{Lai \emph{et~al.}(2015)Lai, Zhao, Law and Adams}]{Lai2015}
Lai ACH, Zhao B, Law AWK, Adams EE. 2015. {A numerical and analytical study of
  the effect of aspect ratio on the behavior of a round thermal}.
  \emph{Environmental Fluid Mechanics} \textbf{15}(1): 85--108,
  \doi{10.1007/s10652-014-9362-3}.

\bibitem[{Langhans \emph{et~al.}(2015)Langhans, Yeo and Romps}]{langhans2015a}
Langhans W, Yeo K, Romps DM. 2015. {Lagrangian Investigation of the
  Precipitation Efficiency of Convective Clouds}. \emph{Journal of the
  Atmospheric Sciences} \textbf{72}(3): 1045--1062,
  \doi{10.1175/JAS-D-14-0159.1}.

\bibitem[{Levine(1959)}]{levine1959}
Levine J. 1959. {Spherical Vortex Theory of Bubble-Like Motion in Cumulus
  Clouds}. \emph{Journal of Meteorology} \textbf{16}(6): 653--662,
  \doi{10.1175/1520-0469(1959)016<0653:SVTOBL>2.0.CO;2}.

\bibitem[{Lilly(1962)}]{lilly1962}
Lilly DK. 1962. {On the numerical simulation of buoyant convection}.
  \emph{Tellus} \textbf{14}(2): 148--172,
  \doi{10.1111/j.2153-3490.1962.tb00128.x}.

\bibitem[{Malkus(1954)}]{malkus1954}
Malkus JS. 1954. {Some Results of a Trade-Cumulus Cloud Investigation}.
  \emph{Journal of Meteorology} \textbf{11}(3): 220--237,
  \doi{10.1175/1520-0469(1954)011<0220:SROATC>2.0.CO;2}.

\bibitem[{Malkus and Scorer(1955)}]{malkus1955a}
Malkus JS, Scorer RS. 1955. {The Erosion of Cumulus Towers}. \emph{Journal of
  the Atmospheric Sciences} \textbf{12}: 43--57,
  \doi{10.1175/1520-0469(1955)012<0000:TEOCT>2.0.CO;2}.

\bibitem[{Mapes and Neale(2011)}]{mapes2011}
Mapes B, Neale R. 2011. {Parameterizing convective organization to escape the
  entrainment dilemma}. \emph{Journal of Advances in Modeling Earth Systems}
  \textbf{3}(2): 1--20, \doi{10.1029/2011MS000042}.

\bibitem[{Miller \emph{et~al.}(1983)Miller, Dye and Martner}]{miller1983}
Miller LJ, Dye JE, Martner BE. 1983. {Dynamical-microphysical evolution of a
  convective storm in a weakly-sheared environment. Part II: Airflow and
  precipitation trajectories from Doppler radar observations}. \emph{Journal of
  the Atmospheric Sciences} \textbf{40}(9): 2097--2109.

\bibitem[{Morrison(2017)}]{morrison2017}
Morrison H. 2017. {An Analytic Description of the Structure and Evolution of
  Growing Deep Cumulus Updrafts}. \emph{Journal of the Atmospheric Sciences}
  \textbf{74}(3): 809--834, \doi{10.1175/JAS-D-16-0234.1}.

\bibitem[{Morrison and Peters(2018)}]{morrison2018}
Morrison H, Peters JM. 2018. {Theoretical Expressions for the Ascent Rate of
  Moist Deep Convective Thermals}. \emph{Journal of the Atmospheric Sciences}
  \textbf{75}(5): 1699--1719, \doi{10.1175/JAS-D-17-0295.1}.

\bibitem[{Morton \emph{et~al.}(1956)Morton, Taylor and Turner}]{morton1956}
Morton BR, Taylor G, Turner JS. 1956. {Turbulent Gravitational Convection from
  Maintained and Instantaneous Sources}. \emph{Proceedings of the Royal Society
  A: Mathematical, Physical and Engineering Sciences} \textbf{234}(1196):
  1--23, \doi{10.1098/rspa.1956.0011}.

\bibitem[{Moser and Lasher-Trapp(2017)}]{moser2017}
Moser DH, Lasher-Trapp S. 2017. {The Influence of Successive Thermals on
  Entrainment and Dilution in a Simulated Cumulus Congestus}. \emph{Journal of
  the Atmospheric Sciences} \textbf{74}(2): 375--392,
  \doi{10.1175/JAS-D-16-0144.1}.

\bibitem[{Murphy \emph{et~al.}(2004)Murphy, Sexton, Barnett, Jones, Webb,
  Collins, Stainforth, Ca, Tuinen, Transport, Overath, Wickner and
  Glatz}]{murphy2004}
Murphy JM, Sexton DM, Barnett DN, Jones GS, Webb MJ, Collins M, Stainforth DA,
  Ca H, Tuinen V, Transport W, Overath P, Wickner B, Glatz A. 2004.
  {Quantification of modelling uncertainties in a large ensemble of climate
  change simulations James}. \emph{Nature} \textbf{430}(August 2004): 768--772,
  \doi{10.1038/nature02771.1.}

\bibitem[{Ogura(1962)}]{ogura1962}
Ogura Y. 1962. {convection of Isolated Masses}.
  \doi{http://dx.doi.org/10.1175/1520-0469(1962)019<0492:COIMOA>2.0.CO;2}.

\bibitem[{Oishi \emph{et~al.}(2018)Oishi, Brown, Burns, Lecoanet and
  Vasil}]{Oishi2018}
Oishi JS, Brown BP, Burns KJ, Lecoanet D, Vasil GM. 2018. {Perspectives on
  Reproducibility and Sustainability of Open-Source Scientific Software from
  Seven Years of the Dedalus Project}. \emph{eprint arXiv:1801.08200} .

\bibitem[{Raymond and Blyth(1986)}]{raymond1986}
Raymond D, Blyth A. 1986. {A stochastic mixing model for nonprecipitating
  cumulus clouds}. \emph{Journal of the atmospheric sciences} \textbf{43}:
  2708--2718, \doi{10.1175/1520-0469(1986)043<2708:ASMMFN>2.0.CO;2}.

\bibitem[{Richards(1961)}]{richards1961}
Richards JM. 1961. {Experiments on the penetration of an interface by buoyant
  thermals}. \emph{Journal of Fluid Mechanics} \textbf{11}(3): 369--384,
  \doi{10.1017/S0022112061000585}.

\bibitem[{Romps(2010)}]{romps2010b}
Romps DM. 2010. {A Direct Measure of Entrainment}. \emph{Journal of the
  Atmospheric Sciences} \textbf{67}(6): 1908--1927,
  \doi{10.1175/2010JAS3371.1}.

\bibitem[{Romps and Charn(2015)}]{romps2015}
Romps DM, Charn AB. 2015. {Sticky Thermals: Evidence for a Dominant Balance
  between Buoyancy and Drag in Cloud Updrafts}. \emph{Journal of the
  Atmospheric Sciences} \textbf{72}(8): 2890--2901,
  \doi{10.1175/JAS-D-15-0042.1}.

\bibitem[{{S{\`a}nchez} \emph{et~al.}(1989){S{\`a}nchez}, {Raymond}, {Libersky}
  and {Petschek}}]{sanchez1989}
{S{\`a}nchez} O, {Raymond} DJ, {Libersky} L, {Petschek} AG. 1989. {The
  Development of Thermals from Rest.} \emph{Journal of Atmospheric Sciences}
  \textbf{46}: 2280--2292,
  \doi{10.1175/1520-0469(1989)046<2280:TDOTFR>2.0.CO;2}.

\bibitem[{Saunders(1961)}]{saunders1961}
Saunders PM. 1961. {An observational study of cumulus}. \emph{Journal of
  Meteorology} \textbf{18}: 451--467.

\bibitem[{Scorer(1957)}]{scorer1957}
Scorer RS. 1957. {Experiments on convection of isolated masses of buoyant
  fluid}. \emph{Journal of Fluid Mechanics} \textbf{2}: 583--594,
  \doi{10.1017/S0022112057000397}.

\bibitem[{Scorer and Ludlam(1953)}]{scorer1953}
Scorer RS, Ludlam FH. 1953. {Bubble theory of penetrative convection}.
  \emph{Quarterly Journal of the Royal Meteorological Society}
  \textbf{79}(339): 94--103, \doi{10.1002/qj.49707933908}.

\bibitem[{Shariff and Leonard(1992)}]{shariff1992}
Shariff K, Leonard A. 1992. {Vortex Rings}. \emph{Annual Review of Fluid
  Mechanics} \textbf{24}(1): 235--279,
  \doi{10.1146/annurev.fl.24.010192.001315}.

\bibitem[{Sherwood \emph{et~al.}(2013)Sherwood, Hern{\'{a}}ndez-Deckers, Colin
  and Robinson}]{sherwood2013}
Sherwood SC, Hern{\'{a}}ndez-Deckers D, Colin M, Robinson F. 2013. {Slippery
  Thermals and the Cumulus Entrainment Paradox*}. \emph{Journal of the
  Atmospheric Sciences} \textbf{70}(8): 2426--2442,
  \doi{10.1175/JAS-D-12-0220.1}.

\bibitem[{Shivamoggi(2010)}]{shivamoggi2010}
Shivamoggi BK. 2010. {Hydrodynamic Impulse in a Compressible Fluid}.
  \emph{eprint arXiv:0910.3223} : 1--10.

\bibitem[{Simpson(1983{\natexlab{a}})}]{simpson1983a}
Simpson J. 1983{\natexlab{a}}. {Cumulus clouds: Early aircraft observations and
  entrainment hypotheses}. In: \emph{Mesoscale Meteorology—Theories,
  Observations and Models.}, Lilly DK, Gal-Chen T\ (eds), Springer: Dordrecht,
  pp. 355--373.

\bibitem[{Simpson(1983{\natexlab{b}})}]{simpson1983b}
Simpson J. 1983{\natexlab{b}}. {Cumulus Clouds: Interactions Between Laboratory
  Experiments and Observations as Foundations for Models}. In: \emph{Mesoscale
  Meteorology — Theories, Observations and Models}, Springer Netherlands:
  Dordrecht, pp. 399--412, \doi{10.1007/978-94-017-2241-4_22}.

\bibitem[{Simpson \emph{et~al.}(1965)Simpson, Simpson, Andrews and
  Eaton}]{simpson1965}
Simpson J, Simpson RH, Andrews DA, Eaton MA. 1965. {Experimental cumulus
  dynamics}. \emph{Reviews of Geophysics} \textbf{3}(3): 387--431,
  \doi{10.1029/RG003i003p00387}.

\bibitem[{Simpson and Wiggert(1969)}]{simpson1969}
Simpson J, Wiggert V. 1969. {Models of Precipitating Cumulus Towers}.
  \emph{Monthly Weather Review} \textbf{97}(7): 471--489,
  \doi{10.1175/1520-0493(1969)097<0471:MOPCT>2.3.CO;2}.

\bibitem[{Squires and Turner(1962)}]{squires1962}
Squires P, Turner JS. 1962. {An entraining jet model for cumulo-nimbus
  updraughts}. \emph{Tellus} \textbf{14}(4): 422--434,
  \doi{10.1111/j.2153-3490.1962.tb01355.x}.

\bibitem[{Stiller and Craig(2001)}]{stiller2001}
Stiller O, Craig GC. 2001. {A scaling hypothesis for moist convective
  updraughts}. \emph{Quarterly Journal of the Royal Meteorological Society}
  \textbf{127}(575): 1551--1570, \doi{10.1256/smsqj.57504}.

\bibitem[{Stirling and Stratton(2012)}]{stirling2012}
Stirling AJ, Stratton RA. 2012. {Entrainment processes in the diurnal cycle of
  deep convection over land}. \emph{Quarterly Journal of the Royal
  Meteorological Society} \textbf{138}(666): 1135--1149, \doi{10.1002/qj.1868}.

\bibitem[{Stommel(1947)}]{stommel1947}
Stommel HM. 1947. {Entrainment of Air Into a Cumulus Cloud}. \emph{Journal of
  Meteorology} \textbf{4}: 91--94,
  \doi{10.1175/1520-0469(1951)008<0127:EOAIAC>2.0.CO;2}.

\bibitem[{Turner(1957)}]{turner1957}
Turner JS. 1957. {Buoyant Vortex Rings}. \emph{Proceedings of the Royal Society
  A: Mathematical, Physical and Engineering Sciences} \textbf{239}(1216):
  61--75, \doi{10.1098/rspa.1957.0022}.

\bibitem[{Turner(1962)}]{turner1962}
Turner JS. 1962. {The ‘starting plume' in neutral surroundings}.
  \emph{Journal of Fluid Mechanics} \textbf{13}(3): 356--368,
  \doi{10.1017/S0022112062000762}.

\bibitem[{Turner(1964{\natexlab{a}})}]{turner1964b}
Turner JS. 1964{\natexlab{a}}. {The dynamics of spheroidal masses of buoyant
  fluid}. \emph{Journal of Fluid Mechanics} \textbf{19}(4): 481--490.

\bibitem[{Turner(1964{\natexlab{b}})}]{turner1964}
Turner JS. 1964{\natexlab{b}}. {The flow into an expanding spherical vortex}.
  \emph{Journal of Fluid Mechanics} \textbf{18}(2): 195--208,
  \doi{10.1017/S0022112064000155}.

\bibitem[{Turner(1986)}]{turner1986}
Turner JS. 1986. {Turbulent entrainment: the development of the entrainment
  assumption, and its application to geophysical flows}. \emph{Journal of Fluid
  Mechanics} \textbf{173}: 431--471.

\bibitem[{Woodward(1959)}]{woodward1959}
Woodward B. 1959. {The motion in and around isolated thermals}. \emph{Quarterly
  Journal of the Royal Meteorological Society} \textbf{85}(August): 144--151.

\bibitem[{Yano(2014)}]{yano2014a}
Yano JI. 2014. {Basic convective element: Bubble or plume? A historical
  review}. \emph{Atmospheric Chemistry and Physics} \textbf{14}(13):
  7019--7030, \doi{10.5194/acp-14-7019-2014}.

\bibitem[{Yeo and Romps(2013)}]{yeo2013}
Yeo K, Romps DM. 2013. {Measurement of Convective Entrainment Using Lagrangian
  Particles}. \emph{Journal of the Atmospheric Sciences} \textbf{70}(1):
  266--277, \doi{10.1175/JAS-D-12-0144.1}.

\bibitem[{Zhao \emph{et~al.}(2013)Zhao, Law, Lai and Adams}]{zhao2013}
Zhao B, Law AW, Lai AC, Adams EE. 2013. {On the internal vorticity and density
  structures of miscible thermals}. \emph{Journal of Fluid Mechanics}
  \textbf{722}: 1--12, \doi{10.1017/jfm.2013.158}.

\bibitem[{Zhao(2014)}]{zhao2014}
Zhao M. 2014. {An investigation of the connections among convection, clouds,
  and climate sensitivity in a global climate model}. \emph{Journal of Climate}
  \textbf{27}(5): 1845--1862, \doi{10.1175/JCLI-D-13-00145.1}.

\bibitem[{Zhao and Austin(2005)}]{zhao2005a}
Zhao M, Austin PH. 2005. {Life Cycle of Numerically Simulated Shallow Cumulus
  Clouds. Part II: Mixing Dynamics}. \emph{Journal of the Atmospheric Sciences}
  \textbf{62}(5): 1291--1310, \doi{10.1175/JAS3415.1}.

\end{thebibliography}

\end{document}